\documentstyle[preprint,eqsecnum,aps,tighten,floats,epsfig]{revtex}

\newcommand{\graphicwidth}{0.8\textwidth}

\newcommand{\ind}{\hspace{.7cm}}

\newcommand{\be}{\begin{equation}}
\newcommand{\ee}{\end{equation}}
\newcommand{\bea}{\begin{eqnarray}}
\newcommand{\eea}{\end{eqnarray}} 
\newcommand{\bml}{\begin{mathletters} \baselineskip 10pt}
\newcommand{\eml}{\baselineskip 12pt \end{mathletters}}
\newcommand{\nn}{\nonumber}

\newcommand{\p}{{\scriptscriptstyle +}}

\newcommand{\bra}{\langle}
\newcommand{\ket}{\rangle}

\newcommand{\cond}{\bra 0 | \bar \psi \psi | 0 \ket}

\newcommand{\e}{\mbox{e}}

\begin{document}

\preprint{\hfill\parbox{4cm}{KYUSHU-HET-39\\ 
                             TPR 97-04\\}}

\draft

\title{Variational   Mass  Perturbation  Theory  for  Light-Front
Bound-State Equations}
\author{Koji Harada\thanks{e-mail: koji@higgs.phys.kyushu-u.ac.jp}}
\address{Department   of  Physics,  Kyushu  University,   Fukuoka
812-81, Japan}
\author{Thomas Heinzl\thanks{e-mail: thomas.heinzl@physik.uni-regensburg.de}, 
Christian Stern\thanks{e-mail: christian.stern@physik.uni-regensburg.de}}
\address{Institut   f\"ur  Theoretische   Physik,   Universit\"at
Regensburg,   Universit\"atsstra\ss   e  31,  93040   Regensburg,
Germany}

\maketitle

\begin{abstract}
\ind We investigate the mesonic light-front bound-state equations
of the 't~Hooft  and Schwinger  model  in the two-particle,  {\em
i.e.}~valence  sector, for small fermion mass.  We perform a high
precision determination  of the mass and light-cone wave function
of the lowest lying meson by combining  fermion mass perturbation
theory with a variational  approach.   All calculations  are done
entirely  in  the  fermionic  representation  without  using  any
bosonization  scheme.  In a step-by-step procedure we enlarge the
space of variational parameters.   We achieve good convergence so
that the calculation of the meson mass squared can be extended to
third order in the fermion mass.  Within a numerical treatment we
include higher Fock states up to six particles.   Our results are
consistent   with  all  previous  numerical  investigations,   in
particular lattice calculations. For the massive Schwinger model,
we find a small discrepancy ($ {\scriptstyle  \lesssim } 2\%$) in
comparison  with known  results.   Possible  resolutions  of this
discrepancy are discussed. 
\end{abstract}


\narrowtext

\section{Introduction}
\label{sec:Introduction}

\ind The number of papers on the Schwinger \cite{Sch62} and 't~Hooft
\cite{tHo74} \cite{tHo75} models is legion -- for a twofold reason.
On the one hand, the two models are particularly simple and become,
for special choices of parameters, even exactly solvable.  On the
other hand, despite their simplicity, both models contain interesting
physics analogous or similar to properties of gauge theories in higher
dimensions.  The models therefore serve as interesting theoretical
laboratories for studying phenomena like confinement and chiral
symmetry breaking, to mention only the most prominent ones.  For
recent reviews on the two models we refer the reader to \cite{Ada97}
for the Schwinger model and to \cite{AA96} for the 't~Hooft model and
generalizations thereof.

\ind
While confinement  (or charge screening)  is realized in the same
way in both models, via a linearly rising Coulomb potential,  the
second feature, chiral symmetry breaking, is not. In the 't~Hooft
model,  1+1 dimensional  QCD in the limit of large $N_C$,  chiral
symmetry  is `almost'  spontaneously  broken  \cite{Wit78}  \cite
{Zhi86}  and a {\em massless}  bound state  arises  in the chiral
limit of vanishing quark mass \cite{tHo74} \cite{tHo75}.   In the
massive   Schwinger   model,   QED  in  $d$  =  1+1  \cite{CJS75}
\cite{Col76},  chiral  symmetry  is anomalously  broken,  and the
contribution  of the anomaly to the mass gap survives  the chiral
limit  yielding  the free,  {\em massive}  boson  of the massless
model.  This boson becomes interacting  in the massive  model and
again can be viewed  as a bound state of fermionic  constituents.
For the sake of brevity  we will call the lowest  bound  state of
both models the `pion' (although, in the real world of $d$ = 3+1,
the    Schwinger     model    boson     corresponds     to    the
$\eta^\prime$-meson).  In this paper we will try to determine its
mass and (light-cone) wave function as accurately as possible. We
mention  in passing  that the Schwinger  model chiral anomaly  is
closely  related  to the  appearance  of a vacuum  $\theta$-angle
parameter \cite{LS71} \cite{KS75} which also affects the particle
spectrum  \cite{Col76}.    Throughout  this  paper,  however,  we
implicitly assume that $\theta$ is set to zero.

\ind
The calculation  of bound state masses and wave functions for the
two models at hand has a long history,  beginning  with the exact
solution  of  the  massless  Schwinger  model  \cite{Sch62}.  For
non-vanishing  `electron'  mass,  $m$,  the  model  is  no longer
exactly solvable and one has to resort to approximations.  One of
them is to assume  the mass $m$ being small and expand around the
massless solution \cite{CJS75} \cite{Col76}.  For the `pion' mass
squared one expects an expansion of the form

\begin{equation}
M^2 (m) = 1 + M_1 m + M_2 m^2 + M_3 m^3 + O(m^4) \; ,
\label{M2EXP}
\end{equation}

where all masses are measured  in units of the basic scale $\mu_0
= e/\sqrt{\pi}$,  the mass of the boson in the massless Schwinger
model, which is thus represented by the `1' in (\ref{M2EXP}). The
first  order  coefficient  $M_1$  was  obtained  analytically  in
\cite{BKS76} via bosonization,

\begin{equation}
M_1 = 2 e^\gamma = 3.56215 \; . 
\label{M_1_EX}
\end{equation}

with  $\gamma  =  0.577216$  being  Euler's  constant.    Shortly
afterwards  Bergknoff,  using light-cone  Hamiltonian  techniques
(see below), found a value \cite{Ber77}

\begin{equation}
M_1 = 2 \pi/\sqrt{3} = 3.62760 \; ,
\label{M_1_BER}
\end{equation}

which differs from (\ref{M_1_EX})  by 1.8 \%.  One topic of the
present  work  will  be the analysis  of this discrepancy  and an
attempt  to  do better  by refining  't~Hooft's  and  Bergknoff's
methods.  This  is  particularly  important  as  the  coefficient
$M_1$  is  directly  related  to  the  vacuum  structure  of  the
Schwinger model by \cite{CHK82}

\begin{equation}
M_1 = - 4\pi \cond \cos \theta \; ,
\label{M_1_COND}
\end{equation}

where $\cond$ is the condensate  of the massless Schwinger  model
\cite{MPR81} \cite{Smi92} (in units of $\mu_0$),

\begin{equation}
\cond = - \frac{1}{2\pi} \e^\gamma = - 0.28347 \; ,
\end{equation}

and $\theta$ the vacuum $\theta$-angle. The result (\ref{M_1_EX})
thus corresponds  to $\theta=0$.   In \cite{Hei96},  the relation
(\ref{M_1_COND})  (for  $\theta=0$),  which  is a 1+1 dimensional
analogue of the Gell-Mann-Oakes-Renner  formula \cite{GOR68}, has
been  used  to determine  the condensate  from  the  `pion'  mass
squared via

\begin{equation}
\cond = \left. - \frac{1}{4\pi} \frac{\partial}{\partial m} M^2
(m) \right\vert_{m=0} = - \frac{1}{4\pi} M_1 \; .
\label{COND}
\end{equation}

Any inaccuracy  in the determination  of $M_1$  thus  immediately
affects the value of the condensate \cite{FP88}. At this point it
should  be mentioned  that within light-cone  quantization  there
have  been  many  attempts  to calculate  the  condensate  of the
massless model alternatively  by solving for its vacuum structure
\cite{LCV}.

\ind
Recently,  the calculations  of $M^2 (m)$ have  been extended  to
second  order.   Using functional  integral  techniques  and mass
perturbation  theory,  Adam  \cite{Ada96}  found  the  analytical
expression

\begin{eqnarray}
M_2 = 4 \pi^2 \cond^2  \, (A \cos 2\theta + B) \, ,
\label{ADAM1} 
\end{eqnarray}

where $A = -0.6599$ and $B = 1.7277$ are numerical constants, given in
terms of integral expressions containing the `pion' propagator for 
$ m = 0 $.  Inserting the values for $A$, $B$ and the condensate, and
setting $\theta=0$, (\ref{ADAM1}) becomes

\begin{equation}
M_2 = 3.3874 \; .
\label{ADAM2}
\end{equation}

This  result  has  been  confirmed  independently  by  Fields  et
al.~\cite{FPV96},  who derived the same integral  expressions  by
summing up all relevant Feynman diagrams in the bosonized  theory
using near-light-cone coordinates.

\ind
To first  order  in $m$, the mass-squared  of the 't~Hooft  model
`pion' has already been calculated  by 't~Hooft  \cite{tHo75}  by
solving  the  associated  light-front  bound-state  equation.  He
derived this equation by projecting  the covariant Bethe-Salpeter
equation onto three-dimensional hypersurfaces of equal light-cone
time, $x^\p$.  Soon afterwards,  the equation was rederived using
light-cone Hamiltonian  techniques \cite{MPT74}.   The result for
the `pion' mass-squared is

\begin{equation}
M^2 (m) = 2 \frac{\pi}{\sqrt{3}} m + O(m^2) \; .
\label{M_1_THO}
\end{equation}

As the light-front  bound  state equations  of the two models  at
hand differ only by an additive  contribution  due to the anomaly
(see  \cite{Hei96}  \cite{BHP90}   and  below),  it  is  not  too
surprising that 't~Hooft's and Bergknoff's results coincide.  For
the 't~Hooft  model,  all masses  are expressed  in units  of the
basic  scale  $\mu_0^2  = g^2  N_C/2\pi$  \footnote{Some  authors
(including 't~Hooft) use a different convention for the coupling,
which amounts to the replacement $g^2 \to 2 g^2$.}. It is obvious
from (\ref{M_1_THO})  that $M^2$ vanishes in the chiral limit, $m
\to 0$. As explained in \cite{Wit78} \cite{Zhi86}, this is not in
contradiction  with Coleman's theorem \cite{Col73}  as the `pion'
is not a Goldstone boson and decouples from the $S$-matrix.

\ind
The condensate of the 't~Hooft model has first been calculated by
Zhitnitsky  using sum rule techniques  \cite{Zhi86}, 

\begin{equation}
\cond/N_C = - \frac{1}{4\pi} M_1 = - 0.28868 \; . 
\label{COND_TH}
\end{equation}

As expected,  the condensate  is proportional  to $N_C$  and  can
again be written in terms of $M_1$. The result has been confirmed
analytically    \cite{Bur89a}    \cite{LLT91}   and   numerically
\cite{Li86}. Via (\ref{COND_TH}), the numerical value \cite{Li86}
for the condensate leads to a numerical estimate for $M_1$,

\begin{equation}
M_1 = 3.64 \pm 0.05 \; .
\label{LI}
\end{equation}

Though  the  numerical   calculation   did  not  use  light-front
methods,   it  seems  to  favour  the  't~Hooft-Bergknoff   value
(\ref{M_1_BER}),  which  is {\it  the}  standard  value  for  the
't~Hooft model.  Higher order corrections to (\ref{M_1_THO}) have
been discussed  in \cite{BSW79}  without explicit calculation  of
the expansion coefficients.

\ind
In recent  years,  both  models  have been  serving  as a testing
ground for new techniques developed in order to solve bound state
problems   using  light-cone   (or,  equivalently,   light-front)
quantization.  These  new  methods  are  `discretized  light-cone
quantization'  (DLCQ) \cite{MY76} \cite{BP85}, where one works in
a finite volume leading to an equally-spaced  momentum  grid, and
the `light-front Tamm-Dancoff (LFTD) approximation' \cite{HPW90},
which  (drastically)  truncates  the Fock space  of the continuum
theory thus limiting the number of constituents a bound state can
have.  The latter approach  can be viewed as a generalization  of
the techniques of 't~Hooft and Bergknoff.   Both methods aim at a
numerical solution of relativistic bound state problems. DLCQ has
been applied to the massive Schwinger  model \cite{BEP87}  and to
QCD  in  1+1  dimensions   (for  arbitrary   $N_C$)  \cite{BHP90}
\cite{Bur89b}.   There are analogous  LFTD calculations  for both
models, QCD$_{1+1}$  \cite{MSY94}  \cite{AI95}  and the Schwinger
model \cite{MP93} \cite{HOT95}.  These latter works are closer in
spirit to ours than the DLCQ approaches.   We will compare to all
this recent work in more detail later on.

\ind
The purpose of this paper is to obtain the `pion' mass squared of
both the Schwinger  and 't~Hooft model to high accuracy including
the third order  in $m$.  In addition,  we want to calculate  the
light-cone  wave function of the `pion' with high precision.   We
will use the light-front techniques of 't~Hooft and Bergknoff and
extensions  thereof.  Our starting point is a LFTD truncation  to
the two-particle sector.  The low order calculations will be done
analytically.   To go beyond,  computer  algebraic  and numerical
methods will be applied and their convergence  tested.  The final
goal is to shed some light on the advantages  and limitations  of
this particular light front approach to bound-state equations.

\ind    The    paper    is    organized     as    follows.     In
Section~\ref{sec:'tHooft's-Ansatz}  we review  't~Hooft's  ansatz
for the wave function  yielding  the lowest order solution of the
bound-state equation. We compare the exact endpoint analysis with
a variational  procedure  which is introduced  at this point.  In
Section~\ref{sec:High-Order-Extension}   we   extend   't~Hooft's
ansatz by adding more variational  parameters  in such a way that
an analytic solution can still be obtained. To this end we employ
computer  algebraic  methods  which  allow  to  treat  up to five
variational parameters. This is sufficient to achieve rather good
convergence.  In Section~\ref{sec:Comp-With-Numer}, these results
are compared  with the outcome of purely numerical  calculations.
We conclude  the paper in Section~\ref{sec:Disc-Concl}  with some
discussion of the presented as well as related work.

\section{'t~Hooft's Ansatz}
\label{sec:'tHooft's-Ansatz}

\ind
Our starting  point is the bound state  equation  of the 't~Hooft
and  Schwinger  model  in the two-particle  sector,  which,  in a
unified way,  can be written as \cite{Hei96} \cite{BHP90} 

\begin{eqnarray}
M^2  \phi(x)  &=&  (m^2  -1)  \frac{\phi(x)}{x(1-x)} -  {\cal P}
\int_0^1 dy \frac{\phi(y)}{(x-y)^2} \nn \\
&+&  \alpha   \int_0^1   dx  \phi(x)   \;   .
\label{BSEQ1} 
\end{eqnarray}

We will refer to this equation as the 't~Hooft-Bergknoff equation
in what follows. $\phi(x)$ denotes the valence part of the `pion'
wave function,  $x$ and $y$ the momentum  fraction  of one of the
two fermions in the meson.  The symbol $ {\cal P}$ indicates that
the  integral  is  defined  as  a  principal  value  \cite{tHo74}
\cite{GS64}.

\ind
The parameter  $\alpha$ measures the strength of the anomaly.  In
the  't~Hooft  model,  $\alpha  = 0$ (no  anomaly),  and  in  the
Schwinger  model  $\alpha  = 1$  (representing the  usual  chiral
anomaly).  The scale parameters,  as already mentioned, are given
by  $\mu_0^2   =  g^2  N_C  /2\pi$   and  $\mu_0^2   =  e^2/\pi$,
respectively.   $M$ denotes  the mass of the lowest  lying  bound
state (the `pion').  Our objective is to obtain (approximate, but
accurate) solutions for $M$ and $\phi$.

\ind
Upon  multiplying  (\ref{BSEQ1})  with $\phi  (x)$ and integrating
over $x$ we obtain for the eigenvalue

\begin{equation}
M^2 (m) = (m^2  - 1) \frac{I_1}{I_0}  - \frac{I_2}{I_0}  + \alpha
\frac{I_3^2}{I_0} \; ,
\label{EIGENVALUE1} 
\end{equation}

where we have defined the integrals

\begin{eqnarray}
\jot4.5pt
I_0 &=& \int_0^1 dx \phi^2 (x) \; , \label{I0} \\ 
I_1 &=& \int_0^1 dx \frac{\phi^2(x)}{x (1-x)} \; , \label{I1} \\
I_2 &=& {\cal P} \int_0^1  dx dy \frac{\phi  (x) \phi (y)}{(x-y)^2}
\; , \label{I2} \\
I_3 &=& \int_0^1 dx \phi (x) \; . \label{I3} 
\end{eqnarray}

$I_0$  is the norm  of the wave  function,  $I_1$  and $I_2$  are
matrix  elements  of  the  mass  and  interaction   term  in  the
light-cone  Hamiltonian  \cite{Ber77}  \cite{MP93}  in the  state
$|\phi  \ket$.   The integral  $I_3$ is the wave function  at the
origin.   An independent  formula for the `pion' mass-squared  is
obtained by integrating (\ref{BSEQ1})  over $x$.  This results in
the simple expression

\begin{equation}
M^2(m) = m^2 \frac{I_4}{I_3} + \alpha \; ,
\label{EIGENVALUE1A}
\end{equation}

with the additional integral,

\begin{equation}
I_4 = \int_0^1 dx \frac{\phi(x)}{x(1-x)} \; .
\label{I4}
\end{equation}

Of course, for the exact wave functions,  the right-hand sides of
(\ref{EIGENVALUE1}) and (\ref{EIGENVALUE1A}) have to coincide. As
the wave functions  cannot be obtained exactly, we will later use
the  agreement  between  both  values  for the mass-squared  as a
measure for the accuracy of our wave functions.  To determine the
latter we will use a particular  class of variational  ans\"atze.

\ind
In his original  work on the subject, \cite{tHo74}  \cite{tHo75},
't~Hooft used the following ansatz for the wave function

\begin{equation}
\phi (x) = x^\beta (1-x)^\beta \, .
\label{THO_ANS}
\end{equation}

This  ansatz  is  symmetric  in $x \leftrightarrow  1-x$  (charge
conjugation odd), and $\beta$ is supposed to lie between zero and
one  so  that  the  endpoint  behaviour  is  non-analytic.  As  a
non-trivial  boundary condition one has the exact solution of the
massless case,

\begin{equation}
M^2  = \alpha  \; , \quad  \mbox{and}  \quad  \phi  (x)  = 1 \; ,
\quad \mbox{{\it i.e.}} \quad \beta = 0 \; . 
\label{MASSLESS} 
\end{equation}

\noindent
The main effect  of having  a non-vanishing  fermion  mass is the
vanishing  of the wave  functions  at the  endpoints  implying  a
non-zero  $\beta$.  This suggests the following  series expansion
for $\beta$,

\begin{equation}
\beta (m) = \beta_1 m + \beta_2 m^2 + \beta_3 m^3 + O(m^4) \; ,
\label{BETA_EXP}
\end{equation}

\noindent
and for the `pion' mass squared,

\begin{equation}
M^2 = \alpha + M_1 m + M_2 m^2 + M_3 m^3 + O(m^4) \; ,
\label{M2_EXP}
\end{equation}

\noindent
in accordance with (\ref{M2EXP}).

\subsection{Exact Endpoint Behaviour}
\label{sec:Exact-Endp-Behav}

\ind
The  exponent   $\beta$  in  (\ref{THO_ANS})   can  actually   be
determined   exactly  by  studying  the  small-$x$  behaviour  of
the bound state equation (\ref{BSEQ1}).   To this end we evaluate
the principal value integral for $x \to 0$,

\begin{eqnarray}
\jot10pt
{\cal P} \int_0^1 dy \, \frac{y^\beta (1-y)^\beta}{(x-y)^2}
&=& x^{\beta-1}        \left[{\cal P}         \int_0^\infty  dz
\frac{z^\beta}{(1-z)^2} + O(x) \right] \nn \\
&=& - x^{\beta-1} \Big[ \pi\beta \cot \pi\beta + O(x) \Big] \; .
\end{eqnarray}

Plugging this into (\ref{BSEQ1}) and demanding the coefficient of
$x^{\beta-1}$  to  vanish  yields  the  transcendental   equation
\cite{tHo74}

\begin{equation}
m^2 - 1 + \pi \beta \cot \pi \beta = 0 \; ,
\label{COT}
\end{equation}

which  is  independent  of  the  anomaly  $\alpha$.   Using  this
expression  we  can  determine  $\beta$  either  numerically  for
arbitrary $m$ or analytically for small $m$ by expanding

\begin{equation}
\beta = \frac{\sqrt{3}}{\pi} m \left(1 - \frac{1}{10} m^2 \right)
+ O(m^4) \; . 
\label{BETA_EXC} 
\end{equation}

Note that the second order coefficient,  $\beta_2$, is vanishing.
Furthermore,  for the exact  $\beta$  one has $\beta_1/\beta_3  =
- 1/10$,  which we will use as a check for our numerical  results
later on.

\ind
Our task is now to determine  the coefficients  $M_i$, $i= 1,2,3$
in  (\ref{M2_EXP}).  The  ansatz  (\ref{THO_ANS})  leads  to  the
following   results   for  the  for  the  `pion'   mass   squared
(\ref{EIGENVALUE1}),

\begin{eqnarray}
M^2 (m, \beta) &=& (m^2 - 1) \Bigg(\frac{1}{\beta} + 4 \Bigg) \nn
\\
&+& \Bigg( \frac{1}{4} + \beta \Bigg) \frac{B^2(\beta , \beta)}{B
(2\beta , 2\beta)} \Bigg[1 + \alpha \frac{\beta}{(2\beta  + 1)^2}
\Bigg] \; . \nn \\
\label{EIGENVALUE2}
\end{eqnarray}

In the above,  $B(z_1  , z_2)$  denotes  the Beta function.   The
relevant formulae for double integrals like $I_2$~(\ref{I2})  can
be  found  in  \cite{BPR80}  and  \cite{HST94},  Appendix~C.  For
$\alpha=0$,   (\ref{EIGENVALUE2})   has  also  been  obtained  in
\cite{AI95}.   Let us emphasize that this result, which expresses
the `pion'  mass  squared  as a function  of the (exactly  known)
endpoint  exponent  $\beta$,  is only  approximate  as 't~Hooft's
ansatz  (\ref{THO_ANS})  for the wave  function  does  {\em  not}
represent  an exact solution of the bound-state  equation.  It is
only  the  endpoint  behaviour  that  is  known  exactly;  in the
intermediate-$x$    region   't~Hooft's   ansatz   is   only   an
approximation that presumably becomes rather bad for large masses
(non-relativistic  limit)  where  the wave  function  is strongly
peaked at $x=1/2$.   The accuracy of the result will be discussed
extensively later on.

\ind
In  order  to  find  $M^2$  to  order  $m^3$  we  need  to expand
(\ref{EIGENVALUE2})  to order $\beta^3$,  as $\beta$ itself is of
order $m$. The result is

\begin{eqnarray}
M^2  (m,  \beta)  &=&  \frac{m^2}{\beta}  +  4  m^2  +  \alpha  +
\frac{\pi^2}{3} \beta \nonumber \\
&+& 4 \Big[  \pi^2/3  - 3 \zeta(3)  + \alpha  (\pi^2/12
- 1) \Big] \beta^2 \nn \\
&+& \Big[ \frac{3}{5}  \pi^4 - 48 \zeta(3)  + 4\alpha  \Big(4 - 3
\zeta(3) \Big) \Big] \beta^3 \nn \\
&+& O(\beta^{\, 4}) \; . 
\label{EIGENVALUE3} 
\end{eqnarray}

Inserting $\beta$ from (\ref{BETA_EXC}) one finds

\begin{eqnarray}
M^2  (m)  &=&  \alpha   +  2 \frac{\pi}{\sqrt{3}} m \nonumber \\
&+& \Bigg[  4 \bigg(2 -  \frac{9}{\pi^2} \zeta(3)  \bigg)  +
\alpha \bigg(1 - \frac{12}{\pi^2}\bigg) \Bigg] m^2 \nonumber \\
&+&  \frac{3\sqrt{3}}{\pi^3} \Bigg[\bigg(  \frac{3}{5}  \pi^4  - 48
\zeta(3) \bigg) + 4 \alpha \Big(4 - 3\zeta(3) \Big) \Bigg] m^3 \nn
\\
&+& O(m^4) \; .
\label{EIGENVALUE4}
\end{eqnarray}

Let us give the explicit  results for the 't~Hooft  and Schwinger
model, {\em i.e.}~for $\alpha$ = 0 and $\alpha$ =1, respectively.
For $\alpha$ = 0, we have

\begin{eqnarray}
M_1 &=&  2\frac{\pi}{\sqrt{3}} = 3.627599 \; , \label{M21_0} \\
M_2 &=& 4 \bigg( 2 - \zeta(3)\frac{9}{\pi^2}  \bigg) = 3.615422
\; , \label{M22_0}\\
M_3 &=& \frac{3\sqrt{3}}{\pi^3} \Big(  \frac{3}{5}  \pi^4  - 48
\zeta(3) \Big) = 0.125139 \; , \label{M23_0} 
\end{eqnarray}

and for $\alpha$ = 1, 

\begin{eqnarray}
M_1 &=&  2\frac{\pi}{\sqrt{3}} = 3.627599 \; , \label{M21_1} \\
M_2  &=&  4  \bigg(  2 - \zeta(3)\frac{9}{\pi^2}  \bigg)  + 1 -
\frac{12}{\pi^2} = 3.399568 \; , \label{M22_1} \\
M_3 &=& \frac{3\sqrt{3}}{\pi^3} \left[ \Big(  \frac{3}{5}  \pi^4  - 48
\zeta(3) \Big) + 4 \Big( 4 - 3 \zeta(3) \Big)\right] \nonumber \\ 
&=& 0.389137 \; . 
\label{M23_1}
\end{eqnarray}

We note that the anomaly  does {\em not} contribute  to the first
order term which therefore is the same in both the Schwinger  and
't~Hooft  model.  This  confirms  the  observation  made  in  the
introduction,    the   coincidence    of   (\ref{M_1_BER})    and
(\ref{M_1_THO}). As $M_1$ is proportional to the condensate, {\it
cf.}~(\ref{COND})   and  (\ref{COND_TH}),   the  latter  is  also
independent of the anomaly \cite{Hei96}.

\ind
For the Schwinger  model ($\alpha$  = 1), the second order result
can  be  compared  with  the  analytical  calculation   of  Adam,
(\ref{ADAM2}).  Astonishingly, the relative difference is smaller
than for the first order, namely 0.36 \%.  This is accidental, as
we shall see below.

\ind As a cross check, we determine the `pion' mass squared using the
alternative formula (\ref{EIGENVALUE1A}).  Plugging in the ansatz
(\ref{THO_ANS}), we find the fairly simple expression

\begin{equation}
\tilde  M^2 (m, \beta) = 4 m^2 + \alpha + \frac{2  m^2}{\beta} \; .
\label{EIGENVALUE5}
\end{equation}

Inserting the expansion (\ref{BETA_EXC}) this becomes

\begin{equation}
\tilde  M^2 = \alpha + \tilde M_1 m + \tilde M_2 m^2 + \tilde
M_3 m^3 + O(m^4) \; ,
\label{EIGENVALUE5A}
\end{equation}

with the coefficients $\tilde M_i$ explicitly given by

\begin{eqnarray}
\tilde M_1 &=&  2 \pi/\sqrt{3} = 3.627599 \; , \\
\tilde M_2 &=& 4 \; , \\
\tilde M_3 &=& \tilde M_1/10 = 0.03627599 \; .
\end{eqnarray}

In (\ref{EIGENVALUE5A})  the anomaly  contributes  only in zeroth
order  (as  $\beta$  is  independent  of $\alpha$),  so that  the
$\tilde  M_i$  are  the  same  for $\alpha=0$  and $\alpha  = 1$.
Comparing with (\ref{M21_0})  and (\ref{M21_1}),  we see that the
first order coefficients  coincide  for the two alternative  mass
formulae.   In addition,  one finds  that  $\tilde  M_2$  roughly
coincides with the $M_2$ of both the 't~Hooft and Schwinger model
to within  10-20 \%.  The value  for $\tilde  M_3$ is smaller  by
approximately an order of magnitude.

\ind
We  conclude   that  't~Hooft's   ansatz  works  well  to  lowest
non-trivial order in $m$ but needs improvement if one wants to go
further.   As a preparative  step  for such a development  we now
introduce  a variational  method that can easily  be extended  to
accurately calculate higher orders in $m$.

\subsection{Variational Approach}
\label{sec:Variational-Approach}

\ind
At variance  with the above we are now going to regard $\beta$ as
a variational  parameter  to  be  determined  by  minimizing  the
function $M^2 (m ,\beta)$ of (\ref{EIGENVALUE3}).  To this end we
insert  the expansion  (\ref{BETA_EXP})  into (\ref{EIGENVALUE3})
and obtain

\begin{eqnarray}
M^2  (m)  &=&  \alpha   +  \Bigg(   \frac{\pi^2}{3}   \beta_1   +
\frac{1}{\beta_1} \Bigg) m \nn \\
&+& \Bigg[ 4 + \beta_2 \Big(\frac{\pi^2}{3} - \frac{1}{\beta_1^2}
\Big) \nn \\ 
&+&  4  \bigg(   \frac{\pi^2}{3}   -  3  \,  \zeta(3)   +  \alpha
\Big(\frac{\pi^2}{12}  -  1 \Big)  \bigg)  \beta_1^2  \Bigg]  m^2
\nonumber \\
&+& \Bigg[ 8 \beta_1 \beta_2  \Big( \frac{\pi^2}{3}  - 3 \zeta(3)
\Big) + \beta_1^3 \Big( \frac{3}{5} \pi^4 - 48 \zeta(3) \Big) \nn
\\
&+& \frac{\beta_2^2}{\beta_1^3}  + \beta_3 \Big(\frac{\pi^2}{3} -
\frac{1}{\beta_1^2} \Big) \nn \\
&+& \alpha \bigg(  2 \beta_1  \beta_2  \Big( \frac{\pi^2}{3}  - 4
\Big) + 4 \beta_1^3  \Big( 4 - 3 \zeta(3) \Big) \bigg) \Bigg] m^3 
\nonumber \\
&+& O(m^4) \; .
\label{EIGENVALUE6}
\end{eqnarray}

\noindent
Note  that to order  $m$ only the leading  coefficient  $\beta_1$
contributes.  We will furthermore shortly see that the dependence
of $M_2$ on $\beta_2$ and of $M_3$ on $\beta_3$ is only apparent.

\ind
Solving the minimization equation, $\partial M^2 / \partial \beta
= 0$, for the coefficients $\beta_i$, leads to

\begin{eqnarray}
\beta_1 &=& \sqrt{3}/\pi = 0.55133 \; , \label{BETA_1_NUM} \\
\beta_2 &=& 0.11690 + 0.065612 \alpha \; , \label{BETA_2_NUM} \\
\beta_3 &=& 0.0049077 - 0.050811 \alpha + 0.019521 \alpha^2 \; . 
\label{BETA_3_NUM} 
\end{eqnarray}

Comparing with (\ref{BETA_EXC}) we note that the coefficient $\beta_1$
is exact!  Plugging it into (\ref{EIGENVALUE6}) we verify the
statement above that $M_2$ is independent of $\beta_2$ and $M_3$
independent of $\beta_3$.  Therefore, the expressions
(\ref{EIGENVALUE6}) and (\ref{EIGENVALUE4}) coincide up to and
including order $m^2$.  $M_1$ and $M_2$ are thus the same,
irrespective of whether one uses the exact endpoint exponent of
(\ref{BETA_EXC}) or its variational estimate.  The estimates
(\ref{BETA_1_NUM}-\ref{BETA_3_NUM}) for the coefficients $\beta_i$
differ in at least three respects from the exact values of
(\ref{BETA_EXC}): (i) $\beta_2$ and $\beta_3$ depend on $\alpha$, (ii)
$\beta_2 \ne 0$, (iii) $\beta_3 \ne -\beta_1/10$.  However, all these
shortcomings affect at most the third order coefficient $M_3$, and
thus become negligible for small $m$.  Only if one wants to have
reliable numbers for $M_3$, (which we will be going to produce), these
effects have to be taken into account.  The present values of
$\beta_i$ lead to

\begin{eqnarray}
M_3 (\alpha=0) &=& 0.043597 \; , \nn \\
M_3 (\alpha=1) &=& 0.190372 \; , 
\end{eqnarray}

which should  be compared  with (\ref{M23_0})  and (\ref{M23_1}),
respectively.  The differences are large so that the agreement is
not particularly  good.  These  discrepancies,  however,  are not
disturbing  at this point,  as we are calculating  a third  order
effect  with just one variational  parameter.   We will do better
later on by enlarging the number of these parameters until we see
satisfactory convergence of the results.

\begin{figure}[tb]
\begin{center}
\caption{Comparison   of  the  two  alternative   mass  formulae,
(\protect\ref{EIGENVALUE2})       for      $M^2(\beta)$       and
(\protect\ref{EIGENVALUE5})  for  $\tilde  M^2(\beta)$,  and  the
second    order   bosonization    results    \protect\cite{Ada96}
\protect\cite{FPV96}  for $m$ = 0.1.  The vertical line marks the
minimum  of  the  curve  $M^2(\beta)$  yielding  the  variational
estimate for $\beta$ using 't~Hooft's  ansatz.  For this value of
$\beta$  one finds $M^2 (m = 0.1) = 1.3969$ and $\tilde  M^2 (m =
0.1)     =    1.3913$.      The    bosonization     result 
(short-dashed horizontal line) is
$M^2_{\mbox{\scriptsize  Bos}}  (m  = 0.1)  = 1.390  \pm  1 \cdot
10^{-3}$, where the error is basically an estimate of the unknown
third order coefficient.   Thus, within  't~Hooft's  ansatz,  the
alternative  mass formula  yields  a somewhat  better  result  at
$m=0.1$.  In the next sections we will refine our methods so that
the  variational  estimates  for  $M^2$  and  $\tilde  M^2$  will
converge towards each other.}
\vspace{0.5cm} 
\includegraphics[width=\graphicwidth]{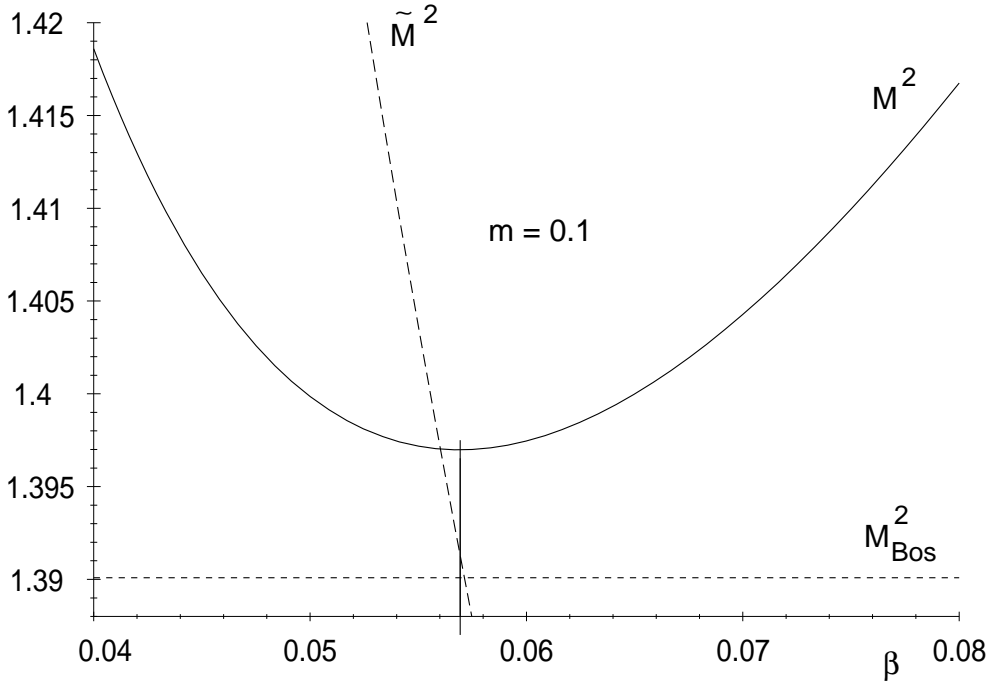}
\end{center} 
\end{figure}

\ind The variational estimate for $\beta$ can also be plugged into the
alternative mass formula, $\tilde M^2 (m, \beta)$,
(\ref{EIGENVALUE5}), which by itself does not constitute a variational
problem.  We do not give the analytical results for the $\tilde M_i$
here but simply refer to Fig.~1 for a qualitative comparison of the
two alternative formulae, and to
Section~\ref{sec:High-Order-Extension}, Tables~\ref{III} and
\ref{VII}, for the actual numbers.

\ind
As stated repeatedly,  't~Hooft's ansatz (\ref{THO_ANS})  used so
far has to be extended if one wants to accurately  determine  the
second  and third order coefficients  of $M^2$.  This is our next
issue.

\section{Extension of 't~Hooft's Ansatz}
\label{sec:High-Order-Extension}

\ind
For  the  numerical  calculations   of  \cite{AI95}  \cite{HOT95}
\cite{HST94} the following set of trial functions has been used:

\begin{equation}
\phi (x) = \sum_{k=0}^N u_k [x(1-x)]^{\beta + k} \; .
\label{THO_EXT1}
\end{equation}

Obviously, the term with $k=0$ and normalization $u_0 = 1$ corresponds
to 't~Hooft's original ansatz (\ref{THO_ANS}).  The coefficients $u_k$
are treated as additional variational parameters, so that, according
to the variational principle, (\ref{THO_EXT1}) {\em must} yield a
better result than (\ref{THO_ANS}).  The question is, how big the
improvement will be. To this end we will calculate the
coefficients $M_i$ by adding more and more basis functions to
't~Hooft's original ansatz (\ref{THO_ANS}), thus enlarging our
space of variational parameters. We will follow the two different
approaches mentioned above, namely, (i) use the exact $\beta$ from
(\ref{BETA_EXC}), or, (ii) treat $\beta$ as one of the variational
parameters.  For both approaches we will compare the calculated
coefficients $M_i$ with the $\tilde M_i$ obtained from the alternative
mass formula (\ref{EIGENVALUE1A}).  We will continue adding basis
functions until we see our results converge.  The maximum number $N$
of basis functions in this section will be five, {\it i.e.}~'t~Hooft's
original  wave  function  plus  four  corrections.  

\ind 
A  particular
benefit  of the ansatz  (\ref{THO_EXT1})  is the  fact  that  the
integrals  $I_1$  to $I_4$  can still  be evaluated  analytically
though  the formulae  become rather lengthy.   For this reason we
will  make  heavy  use  of  the  program  package  MAPLE  in what
follows.  As a result we have analytical expressions  for all the
quantities  we calculate.   As these expressions  cover pages and
pages without being very instructive,  we do not display them but
rather the evaluated numbers.  In this sense, our treatment might
be called `semi-analytical'. A major advantage, however, of using
computer  algebraic  manipulations   is  the  high  calculational
accuracy  which is only limited  by the maximum  number of digits
the machine can handle.

\subsection{Purely Variational Approach}
\label{sec:Purely-Vari-Appr}

\ind
In  this  subsection,   $\beta$  will  always  be  treated  as  a
variational  parameter  and thus  be obtained  by minimizing  the
`pion' mass squared, $M^2$. In the ansatz (\ref{THO_EXT1}) we use
up to four additional basis functions, so that our maximum $N$ is
four. To avoid an inflation of indices we rename the coefficients
$u_k$, $k= 1, \ldots , 4$ by $a$, $b$, $c$ and $d$, respectively.
Their expansion coefficients, defined via

\begin{eqnarray}
    a   &=&   a_1   m +   a_2   m^2  \, , \nn \\
    b   &=&   b_1   m +   b_2   m^2  \, , \nn \\
    c   &=&   c_1   m +   c_2   m^2  \, , \nn \\
    d   &=&   d_1   m +   d_2   m^2  \, ,
\label{COEFFS}    
\end{eqnarray}

are determined (recursively)  by minimizing $M^2$ with respect to
them. Let us start with the 't~Hooft model ($\alpha=0$).

\subsubsection{'t~Hooft Model}
\label{sec:'tHooft-Model}

\ind
In order to save space we do not display  all the values  for the
coefficients  in (\ref{COEFFS}).   The best values  obtained  are
given  in the last  section  when we discuss  the quality  of the
wave  function.   Here,  we rather  display  the results  for the
coefficients  $\beta_i$ of $\beta$ (see Table \ref{I}), as these can be
compared with the exact values of (\ref{BETA_EXC}).

\ind
As an important  finding we note that $\beta_1$ remains unchanged
at the exact  value $\sqrt{3}/\pi$  (a `variational  invariant').
$\beta_2$  tends  to zero,  as it should.   The non-vanishing  of
$\beta_2$ only affects the coefficient  $M_3$, as $M_1$ and $M_2$
do not depend on $\beta_2$.  The convergence of $\beta_3$ towards
$-\beta_1/10$  seems somewhat  slow; but as all $M_i$, $i = 1, 2,
3$, are independent of $\beta_3$ this has no observable effect.

\ind
In Table~\ref{II}  we list the expansion coefficients of the mass
squared, $M^2$. One notes that the convergence of the results for
$M_2$  and $M_3$  is rather  good.   For $M_2$ we finally  have a
relative accuracy of $8 \cdot 10^{-7}$, and for $M_3$ of $4 \cdot
10^{-5}$.  Furthermore,  the coefficients  are getting smaller if
one adds more basis functions, in accordance with the variational
principle. As the coefficient $M_1$ is entirely determined by the
`variational   invariant'   $\beta_1$  it  remains  unchanged  at
$2\pi/\sqrt{3}$.   This is another important result:  $M_1$ stays
fixed at the standard 't~Hooft-Bergknoff value (\ref{M_1_BER}).

\ind
If we use the alternative  mass formula (\ref{EIGENVALUE1A})  and
evaluate it using the wave function calculated  above we find the
values for $\tilde{M}^2$ listed in Table \ref{III}.

\ind
It should be pointed out that the values of the $\tilde  M_i$ are
somewhat   more  sensitive   to  the  values   of  the  expansion
coefficients  $\beta_i$,  because  $\tilde  M_2$  does depend  on
$\beta_2$, and $\tilde M_3$ on $\beta_2$ and $\beta_3$.  To check
the quality of our wave functions  one should compare the results
for $ M_i $ and $ \tilde M_i$ which are listed in Tables~\ref{II}
and~\ref{III}.   Once  more it is gratifying  to note that we are
improving our results step by step in the variational  procedure.
The values of $\tilde M_i$ converge  towards those of $M_i$.  The
relative  accuracy  is approximately  $10^{-3}$.   This error  is
entirely  due to the difference  between  the calculated  and the
real wave function.   It is clear that the accuracy in the single
eigenvalue $M^2$ (Table~\ref{II})  is higher than the one for the
wave function,  where in principle  an infinite  number of points
have to be calculated.  This represents an indirect proof that we
are  indeed  improving  our  wave  functions,  and  not just  the
eigenvalues.

\begin{table}
\renewcommand{\arraystretch}{1.5}
\caption{\label{I}Expansion   coefficients   of  the  end   point
exponent  $\beta$ for the 't~Hooft model obtained by successively
varying  with respect  to $\beta$  (first line) , $\beta$ and $a$
(second line), etc.  The first column, $\beta_1$,  coincides with
the result  from the exact endpoint  analysis,  which in addition
yields $\beta_2 = 0$, $\beta_3 = - \beta_1/10$.} 
\vspace{.3cm}

\begin{tabular}[t]{l l d d }
  \it Ansatz & $\beta_1 = \sqrt{3}/\pi$ & $\beta_2$ & $\beta_3$ \\
  \hline
  't~Hooft & 0.55132889 & 0.11689763  &    0.00490773  \\ 
  a        & 0.55132889 & 0.00976951  & $-$0.04010634   \\ 
  b        & 0.55132889 & 0.00256081  & $-$0.04889172   \\ 
  c        & 0.55132889 & 0.00102086  & $-$0.05209341   \\ 
  d        & 0.55132889 & 0.00050738  & $-$0.05343097   \\ 
\end{tabular}
\end{table}

\begin{table}
\renewcommand{\arraystretch}{1.5}
\caption{\label{II}Expansion   coefficients   of  $M^2$  for  the
't~Hooft  model obtained by successively  enlarging  the space of
variational parameters.  $M_1$ is the standard 't~Hooft-Bergknoff
result.   Note  the good convergence  towards  the bottom  of the
table.} 
\vspace{.3cm}

\begin{tabular}[t]{l c c c }
  \it Ansatz & $M_1 = 2\pi/\sqrt{3}$ & $M_2$ & $M_3$ \\
  \hline
  't~Hooft & 3.62759873 & 3.61542218 & 0.043597197  \\ 
  a        & 3.62759873 & 3.58136872 & 0.061736701  \\ 
  b        & 3.62759873 & 3.58107780 & 0.061805257  \\ 
  c        & 3.62759873 & 3.58105821 & 0.061795547  \\ 
  d        & 3.62759873 & 3.58105532 & 0.061793082  \\ 
\end{tabular}
\end{table}

\begin{table}
\renewcommand{\arraystretch}{1.5}
\caption{\label{III}Expansion  coefficients  of  the  alternative
mass squared  $\tilde  M^2$ for the 't~Hooft  model  obtained  by
successively enlarging the space of variational parameters. Again,
the   fixed   value   for   $\tilde    M_1$   is   the   standard
't~Hooft-Bergknoff result.} 
\vspace{.3cm}

\begin{tabular}[t]{l c c c }
  \it Ansatz & $\tilde{M}_1 = 2\pi/\sqrt{3}$ & $\tilde{M}_2$ 
  & $\tilde{M}_3$ \\
  \hline
  't~Hooft & 3.62759873 & 3.23084437 & 0.130791594   \\ 
  a        & 3.62759873 & 3.54922830 & 0.066592425  \\ 
  b        & 3.62759873 & 3.57265307 & 0.059652468  \\ 
  c        & 3.62759873 & 3.57769969 & 0.060282959  \\ 
  d        & 3.62759873 & 3.57938609 & 0.060854772   \\ 
\end{tabular}
\end{table}

\subsubsection{Schwinger Model}
\label{sec:Schwinger-Model}

\ind
For the Schwinger  model ($\alpha$ = 1), we perform  exactly  the
same calculations.   The expansion  coefficients  of $\beta$  are
listed in Table~\ref{V}.
Although  the  coefficients  $\beta_2$,  $\beta_3$  are  somewhat
different  from  those  in Table \ref{I} (which  they  must be as they
lose their dependence  on $\alpha$  only in the strict  limit  of
{\it  exact}  evaluation)  we note the same tendency:   $\beta_2$
converges to zero and $\beta_3$ towards $-\beta_1/10$.

\begin{table}
\renewcommand{\arraystretch}{1.5}
\caption{\label{V}   Expansion   coefficients   of  the  endpoint
exponent $\beta$ for the Schwinger model obtained by successively
enlarging the space of variational parameters.  The first column,
$\beta_1$,  coincides  with  the result  from the exact  endpoint
analysis,  which in addition yields $\beta_2  = 0$, $\beta_3  = -
\beta_1/10$.} 
\vspace{.3cm}

\begin{tabular}[t]{l c c c }
  \it Ansatz & $\beta_1$ & $\beta_2$ & $\beta_3$ \\
  \hline
  't~Hooft & 0.55132889 & 0.18250945 & $-$0.026382222  \\ 
  a        & 0.55132889 & 0.01177681 & $-$0.040372413  \\ 
  b        & 0.55132889 & 0.00317437 & $-$0.048081756  \\ 
  c        & 0.55132889 & 0.00127507 & $-$0.051602713  \\ 
  d        & 0.55132889 & 0.00063609 & $-$0.053130398  \\ 
\end{tabular}
\end{table}

\begin{table}
\renewcommand{\arraystretch}{1.5}
\caption{\label{VI}Expansion  coefficients of $M^2$ for the Schwinger model
  obtained by successively enlarging the space of variational
  parameters.  Again, the fixed value for $M_1$ is the standard
  't~Hooft-Bergknoff result.  Note the good convergence towards the
  bottom of the table.}

\vspace{.3cm}
\begin{tabular}[t]{l c c c }
  \it Ansatz & $M_1 = 2\pi/\sqrt{3}$ & $M_2$ & $M_3$ \\
  \hline
  't~Hooft & 3.62759873 & 3.39956798 & 0.19037224  \\ 
  a        & 3.62759873 & 3.30906326 & 0.34776772  \\ 
  b        & 3.62759873 & 3.30864244 & 0.34820193  \\ 
  c        & 3.62759873 & 3.30861240 & 0.34820513  \\ 
  d        & 3.62759873 & 3.30860791 & 0.34820389  \\ 
\end{tabular}
\end{table}

\begin{table}
\renewcommand{\arraystretch}{1.5}
\caption{\label{VII}  Expansion  coefficients  of the alternative
mass squared  $\tilde  M^2$ for the Schwinger  model obtained  by
successively  enlarging  the  space  of  variational  parameters.
Again,  the  fixed  value  for  $\tilde   M_1$  is  the  standard
't~Hooft-Bergknoff result.}

\vspace{.3cm}  
\begin{tabular}[t]{l  c  c  c  } \it  Ansatz  & $\tilde{M}_1  =
2\pi/\sqrt{3}$ & $\tilde{M}_2$ & $\tilde{M}_3$ \\ \hline
't~Hooft  &  3.62759873  &  2.79913596  & 0.57111672  \\  
a         &  3.62759873  &  3.27031909  & 0.36868190  \\ 
b         &  3.62759873  &  3.29819920  & 0.34869835  \\ 
c         &  3.62759873  &  3.30441759  & 0.34753576  \\ 
d         &  3.62759873  &  3.30651525  & 0.34762562  \\
\end{tabular} 
\end{table}

\ind
The best values for the `pion' mass squared are again provided by
the  variational  results  listed  in  Table~\ref{VI}.   The  numerical
accuracy is practically  the same as for the analogous  Table~\ref{II}.
The comparison  with  the alternatively  calculated  coefficients
$\tilde M^2$ is given in Table~\ref{VII}.
Again,  everything  is completely  analogous  to the case  of the
't~Hooft model ($\alpha=0$).

\subsection{Variational Approach using the exact $\beta$}
\label{sec:Vari-Appr-using}

\ind In this subsection we calculate $M^2$ and $\tilde M^2$ for both
values of $\alpha$ using the exact value (\ref{BETA_EXC}) for $\beta$.
Thus only $a$, $b$, $c$, $d$ are treated as variational parameters.
This is the procedure employed numerically in \cite{HOT95}. From the
discussion of the preceding subsection, in particular the values for
the $\beta_i$ displayed in Table~\ref{I}, which differ minimally from
the exact values, we expect that the results will be very close to
those from the purely variational approach.

\subsubsection{'t~Hooft Model}
\label{sec:'tHooft-Model-2}

\ind This is exactly what happens as can be seen from Tables~\ref{IX}
and \ref{X}.  Comparing Table~\ref{IX} with Table~\ref{II} one finds
that the coefficients  $M_1$ and $M_2$ of both tables coincide as
these  only  depend  on  $\beta_1$  which  is  the  same  in both
approaches.  Only for $M_3$ there are slight differences,  due to
the dependence on $\beta_2$.   For the $\tilde M_i$, $i=2,3$, the
discrepancies are somewhat bigger as these coefficients do depend
on    $\beta_2$    and    $\beta_3$     (Table~\ref{III}     {\em
vs.}~Table~\ref{X}). Still the consistency is quite obvious.

\ind Upon comparing the second columns of Tables~\ref{II} and
\ref{III}, respectively Tables~\ref{IX} and \ref{X}, one notes a funny
coincidence. In the first case $M_2 $ is slightly bigger than $\tilde
M_2$, and vice versa in the second case.  If one denotes the purely
variational results with a superscript `$v$', and the results obtained
with the exact $\beta$ with a superscript `$e$', one finds that $M_2$,
which is the same in both approaches, is given by the arithmetic mean

\begin{equation}
M_2 = \frac{1}{2} \left(\tilde M_2^e + \tilde M_2^v \right) \; .
\end{equation}

We have checked this analytically for 't~Hooft's ansatz.  For the
higher orders this is difficult  to do but the numerical evidence
is beyond doubt.

\ind Comparing the third columns of Tables~\ref{II} and \ref{IX}, one
finds that $ M_3^v < M_3^e $.  We thus see a slight tendency that the
results of the purely variational procedure are better than those
obtained using the exact $\beta$.  The same observation has been made
by Mo and Perry \cite{MP93}, in particular for large values of the
fermion mass $m$ (where the third order coefficient $M_3$ becomes
important).  We interpret this fact as a hint that for larger $m$ the
behaviour of the wave function in the intermediate region becomes more
relevant compared to the endpoint behaviour.

\begin{table}
\renewcommand{\arraystretch}{1.5}
\caption{\label{IX}  Expansion  coefficients  of $M^2$  ('t~Hooft
model) obtained by using the exact endpoint exponent  $\beta$ and
successively  enlarging  the  space  of variational  parameters..
Again,    the   fixed   value   for   $M_1$   is   the   standard
't~Hooft-Bergknoff result.}

\vspace{.3cm}

\begin{tabular}[t]{l c c c } \it Ansatz & $M_1 = 2\pi/\sqrt{3}$
& $M_2$ & $M_3$ \\ \hline
't~Hooft & 3.62759873  & 3.61542219  & 0.12513878 \\
a        & 3.62759873  & 3.58136873  & 0.06230622 \\ 
b        & 3.62759873  & 3.58107781  & 0.06184441 \\
c        & 3.62759873  & 3.58105821  & 0.06180178  \\ 
d        & 3.62759873  & 3.58105533  & 0.06179462 \\ 
\end{tabular} 
\end{table}

\begin{table}
\renewcommand{\arraystretch}{1.5}
\caption{\label{X}  Expansion  coefficients  of the alternatively
defined  mass squared $\tilde M^2$ ('t~Hooft  model) obtained  by
using  the  exact  endpoint  exponent  $\beta$  and  successively
enlarging the space of variational  parameters.  Again, the fixed
value  for  $\tilde  M_1$  is  the  standard   't~Hooft-Bergknoff
result.}

\vspace{.3cm}

\begin{tabular}[t]{l l l l }
  \it Ansatz & $\tilde{M}_1 = 2\pi/\sqrt{3}$ 
  & $\tilde{M}_2$ & $\tilde{M}_3$ \\
  \hline
  't~Hooft & 3.62759873 & 4.0        & 0.36275987   \\ 
         a & 3.62759873 & 3.61350915 & 0.08659575  \\ 
         b & 3.62759873 & 3.58950254 & 0.07106398  \\ 
         c & 3.62759873 & 3.58441672 & 0.06603497  \\ 
         d & 3.62759873 & 3.58272458 & 0.06406179  \\ 
\end{tabular}
\end{table}

\subsubsection{Schwinger Model}
\label{sec:Schwinger-Model-2}

\ind For the Schwinger model, the analogous results are listed in
Tables \ref{XII} and \ref{XIII}.  Exactly the same remarks as above
apply including the size of the errors and the relation between $M_2$
and $\tilde M_2^e$, $\tilde M_2^v$.  In the next section we will
verify our results by purely numerical methods.

\begin{table}
\renewcommand{\arraystretch}{1.5}
\caption{\label{XII}  Expansion coefficients  of $M^2$ (Schwinger 
model) obtained by using the exact endpoint exponent  $\beta$ and
successively  enlarging  the  space  of  variational  parameters. 
Again,    the   fixed   value   for   $M_1$   is   the   standard
't~Hooft-Bergknoff result.}

\vspace{.3cm}
\begin{tabular}[t]{l c c c }
  \it Ansatz & $M_1 = 2\pi/\sqrt{3}$ & $M_2$ & $M_3$ \\
  \hline
  't~Hooft & 3.62759873 & 3.39956798 & 0.38913656  \\ 
         a & 3.62759873 & 3.30906326 & 0.34859533  \\ 
         b & 3.62759873 & 3.30864244 & 0.34826210  \\ 
         c & 3.62759873 & 3.30861239 & 0.34821485  \\ 
         d & 3.62759873 & 3.30860791 & 0.34820630   \\ 
\end{tabular}
\end{table}

\begin{table}
\renewcommand{\arraystretch}{1.5}
\caption{\label{XIII} Expansion coefficients of the alternatively
defined mass squared $\tilde M^2$ (Schwinger  model) obtained  by
using  the  exact  endpoint  exponent  $\beta$  and  successively
enlarging the space of variational  parameters.  Again, the fixed
value  for  $\tilde  M_1$  is  the  standard   't~Hooft-Bergknoff
result.}

\vspace{.3cm}
\begin{tabular}[t]{l l l l }
  \it Ansatz & $\tilde{M}_1 = 2\pi/\sqrt{3}$ 
  & $\tilde{M}_2$ & $\tilde{M}_3$ \\
  \hline
  \hline
  't~Hooft & 3.627598730 & 4.0         & 0.3627598730  \\ 
         a & 3.627598730 & 3.347807427 & 0.3637551073  \\ 
         b & 3.627598730 & 3.319085690 & 0.3566261465  \\ 
         c & 3.627598730 & 3.312807215 & 0.3523062162  \\ 
         d & 3.627598730 & 3.310700593 & 0.3504584408   \\ 
\end{tabular}
\end{table}

\section{Comparison With Numerical Results}
\label{sec:Comp-With-Numer}

\ind
For the numerical  calculations  the ansatz (\ref{THO_EXT1})  has
been used.  The value for $\beta$ has been determined numerically
from (\ref{COT}).   Again, up to four additional  basis functions
have been included  (for larger  masses even five).  Let us begin
with the Schwinger model ($\alpha=1$).   Here we can use the code
developed  in \cite{HOT95}.   In Table~\ref{XV}  we list $M^2  -1$ as a
function of the fermion mass $m$, calculated  within two-particle
LFTD approximation. The notations a,b,c,d are as in the preceding
section,  the letter `e' denotes  the inclusion  of a fifth basis
function.

{\squeezetable
\widetext

\begin{table}
\renewcommand{\arraystretch}{1.5}
\caption{\label{XV}   Numerical   results   using  lowest   order
(two-particle)  LFTD approximation  for $M^2 -1$ as a function of
the fermion mass $m$, for $\alpha = 1$ (Schwinger model).}

\vspace{.3cm}
\begin{tabular}[t]{l c c c c c c c c c }
\it Ansatz & $m$ = 0.0001 & $m$ = 0.0005 & $m$ = 0.001 & $m$ = 0.005 & $m$ = 0.01 & $m$ = 0.05 & $m$ = 0.1 & $m$=0.3 & $m$=0.5 \\ 
\hline  
't~Hooft  & 0.000362794 & 0.00181465 & 0.00363100 & 0.0182230 & 0.0366163 & 0.189927 & 0.397130 & 1.403677 & 2.705008 \\
a         & 0.000362858 & 0.00181465 & 0.00363086 & 0.0182207 & 0.0366072 & 0.189695 & 0.396181 & 1.394169 & 2.675165 \\
b         & $-$         & 0.00181461 & 0.00363100 & 0.0182210 & 0.0366071 & 0.189694 & 0.396176 & 1.394129 & 2.675072 \\
c         & $-$         & 0.00181459 & $-$        & $-$       & 0.0366071 & 0.189694 & 0.396176 & 1.394125 & 2.675059 \\
d         & $-$         & $-$        & $-$        & $-$       & 0.0366073 & 0.189694 & 0.396176 & 1.394125 & 2.675057 \\
e         & $-$         & $-$        & $-$        & $-$       & $-$       & $-$      & $-$      & 1.394123 & 2.675058 \\
\end{tabular}
\end{table}
\narrowtext
}

\ind One main difference in comparison with the computer algebraic
treatment of Section~\ref{sec:High-Order-Extension} is the numerical
inaccuracy for small $m$.  This is a general disease of numerical
treatments, and also shared {\it e.g.}~by the lattice \cite{CHK82},
\cite{MPR81} or DLCQ approach \cite{BEP87}.  In these approaches,
however, the numerical errors are typically much larger than ours (see
the next section for an explicit comparison).  Due to the small-$m$
instability, our numerical calculation does not converge within an
arbitrary number of digits.  As soon as a calculated value for $M^2 -
1$ becomes bigger as the preceding one (a numerical violation of the
variational principle), we terminate the procedure and pick the
smaller value as our final result.  The difference between these last
two values can be used as an estimate of the numerical inaccuracy.
The errors will be further discussed below and in the next section
when we compare our results with related work (see Tables~\ref{XIX},
\ref{XX}).

\ind
It should  also be pointed  out that  the numerical  analysis  is
conceptually   very   different   from  the  computer   algebraic
treatment.  As one cannot perform a Taylor expansion numerically,
the  expansion  coefficients  of  $M^2$  have  to be obtained  by
fitting polynomials to $M^2 (m)$.  Clearly, this is an additional
source of errors, and one expects the results to be less accurate
than those of the preceding sections.  A cubic fit to the optimum
values  for $M^2 -1$ in Table~\ref{XV}  yields the following  expansion
coefficients of $M^2$,

\bml
\label{EVNUM1_3}
\begin{equation}
M_1 =  3.62609 \; , 
\end{equation}
\begin{equation}
M_2 =  3.33607 \; , 
\end{equation}
\begin{equation}
M_3 =  0.22396 \; ,
\end{equation}
\eml

which  should  be compared  with the last  line of Table~\ref{VI}.   To
estimate  the accuracy  of these values  we also show the results
from a quartic fit, 

\bml
\label{EVNUM1_4}
\begin{equation}
M_1 =  3.62755 \; , 
\end{equation}
\begin{equation}
M_2 =  3.31029 \; , 
\end{equation}
\begin{equation}
M_3 =  0.32984 \; .
\end{equation}
\eml

Comparing  (\ref{EVNUM1_3})  and (\ref{EVNUM1_4})  one finds that
the stability of the fits is not too impressive in view of the
accuracy we would like to achieve.  In particular the third order
coefficient  is numerically  difficult to determine.  We estimate
the relative accuracy as being $10^{-3}$ for $M_1$, $10^{-2}$ for
$M_2$  and  $10^{-1}$  for  $M_3$.  This  is  also  confirmed  by
comparing with Table~\ref{VI}.

\ind
For the Schwinger  model it is possible to check the influence of
higher  particle  sectors  on the `pion'  mass squared  by using  the
machinery developed in \cite{HOT95} for the wave functions of the
higher Fock components.   These are the amplitudes of finding not
only two, but four, six, ...  (anti-)fermions  in the `pion'.  In
Table~\ref{XVI},  we list the best values for $M^2-1$  including  up to
six-body  wave  functions.   We also show  the 2-, 4-, and 6-body
content of the total wave function.   It is known that the `pion'
is entirely 2-particle in the chiral limit \cite{MP93}. For small
mass,  one therefore  expects  only small contributions  from the
higher Fock sectors.  This is confirmed by the numerical results.
Astonishingly,  this feature  persists  up to values  of at least
$m=0.5$ for the fermion mass.  This fact is in agreement with the
observation  of Mo and Perry that the four-particle  component of
the wave function is less than 0.4~\% for {\em all} values of the
fermion mass \cite{MP93}.  A similar result has been found in the
DLCQ calculations of \cite{BEP87}. In addition we note that there
seems  to  be some  interesting  kind  of hierarchy  between  the
relative strengths of the contributions  from different  particle
sectors.   If we denote the $2k$-particle  amplitude  in the wave
function  by $f_{2k}$,  we find  that  $|f_2|^2  \gg |f_4|^2  \gg
|f_6|^2$,  the individual  proportions  being several  orders  of
magnitude  (see Table~\ref{XVI}).   Here,  we explicitly  see the
magic of light-front  field theory at work:  high Fock components
in bound states tend to be largely suppressed,  at variance  with
the situation encountered  in field theory quantized the standard
way.

{\squeezetable
\widetext
\begin{table}
\renewcommand{\arraystretch}{1.5}
\caption{\label{XVI} Numerical Results for $M^2 -1$ as a function
of the fermion  mass  $m$,  for  $\alpha  = 1$ (Schwinger  model)
including 2-, 4- and 6-body wave functions.  The contributions of
the different Fock sectors to the total wave function squared are
given in percent.}

\vspace{.3cm}
\begin{tabular}[t]{l c c c c c c c c }
  \it       & $m$ = 0.0001      & $m$ = 0.0005     & $m$ = 0.001      & $m$ = 0.005     & $m$= 0.01 & $m$ = 0.05 & $m$ = 0.1 & $m$ = 0.5 \\   \hline
  $M^2 - 1$ & 0.000362793 & 0.00181481 & 0.00363055 & 0.0182179 & 0.0365968 & 0.189468  & 0.395400  & 2.66787 \\ 
  \% 2-body & 100.00      & 100.00     & 100.00     & 99.999990 & 99.999962 & 99.999213 & 99.997514 & 99.986816 \\
  \% 4-body & $-$         & $-$        & $-$        & 0.000010  & 0.000038  & 0.000782  & 0.002474  & 0.013171 \\
  \% 6-body & $-$         & $-$        & $-$        & $-$       & $-$       & 0.000005  & 0.000013  & 0.000013 \\ 
  \end{tabular} 
  \end{table}

\narrowtext
}

\ind
From the point of view of the variational  principle, the results
shown  in Table~\ref{XVI}  are a bit better  ({\em i.e.}~smaller)  than
those of Table~\ref{XV} (apart from the value for $m=0.0005$),  but the
improvement is rather small. A cubic fit yields

\bml
\label{EVNUM2_3}
\begin{equation}
M_1 =  3.62667 \; , 
\end{equation}
\begin{equation}
M_2 =  3.23696 \; , 
\end{equation}
\begin{equation}
M_3 =  0.36235 \; ,
\end{equation}
\eml

and a quartic fit

\bml
\label{EVNUM2_4}
\begin{equation}
M_1 =  3.62747 \; , 
\end{equation}
\begin{equation}
M_2 =  3.20864 \; , 
\end{equation}
\begin{equation}
M_3 =  0.60365 \; .
\end{equation}
\eml

\ind
Compared  with  (\ref{EVNUM1_3})  and (\ref{EVNUM1_4})  we do not
find   any  absolute   improvement   in  $M_1$   that  could   be
distinguished from the numerical inaccuracy.  For the coefficient
$M_2$, which in bosonization  schemes is 3.3874,  we even get the
wrong tendency:   it becomes  smaller upon including  higher Fock
states.   The  inaccuracy   for  $M_3$  is  so  large  that  this
coefficient is only determined in its order of magnitude.

\ind
Altogether,  we arrive at the very important conclusion  that the
inclusion  of higher particle  sectors  in the light-front  bound
state  equation  of the Schwinger  model  does  not diminish  the
discrepancy    between    the    results    obtained    via   the
't~Hooft-Bergknoff method and those from bosonization techniques.

\ind
Let us move to the 't~Hooft  model.  If we put $\alpha=0$  in the
Schwinger  model  code  above  (Code~I)  we find the `pion'  mass
squared  for the 't~Hooft  model.   The best  values  within  the
2-particle  sector  are listed  in Table~\ref{XVII}.   Note that higher
particle  sectors  are  strictly  suppressed  in the  large-$N_C$
limit.  Thus, there is no point in calculating  these, unless one
is  interested  in  $1/N_C$  corrections  to the  't~Hooft  model
results \cite{BHP90} \cite{Bur89b} \cite{MSY94}.

\begin{table}
\renewcommand{\arraystretch}{1.5}
\caption{\label{XVII}  $M^2$  as a function  of the fermion  mass
$m$, for $\alpha  = 0$ ('t~Hooft  model),  obtained  with Code I}
\vspace{.3cm}

\begin{tabular}[t]{l l l l l }
  \it $m$ & 0.0001 & 0.001 & 0.01 & 0.1 \\
  \hline
  $  M^2$  & 0.000362795 & 0.00363109 & 0.0366342 & 0.398634 \\ 
\end{tabular}
\end{table}

\ind
In view of the restricted  number of data points only a quadratic
fit makes sense which yields 

\bml
\label{EVNUM3}
\begin{equation}
M_1 =  3.62754 \; , 
\end{equation}
\begin{equation}
M_2 =  3.58796 \; ,
\end{equation}
\eml

which is consistent with the results of
Section~\ref{sec:High-Order-Extension} (see Table~\ref{II}).  A second
code (Code II), which was independently developed for the 't~Hooft
model \cite{MSY94} yields the results displayed in Table~\ref{XVIII}.
From a quadratic fit we obtain

\begin{table}
\renewcommand{\arraystretch}{1.5}
\caption{\label{XVIII}  $M^2$ as a function  of the fermion  mass
$m$, for $\alpha = 0$ ('t~Hooft model), obtained with Code II} 
\vspace{.3cm}

\begin{tabular}[t]{l l l l l }
  \it $m$ & 0.0001 & 0.001 & 0.01 & 0.1 \\
  \hline
  $  M^2$  & 0.000362795 & 0.00363118 & 0.0366341 & 0.398634 \\ 
\end{tabular}
\end{table}

\bml
\label{EVNUM4}
\begin{equation}
M_1 =  3.62754 \; , 
\end{equation}
\begin{equation}
M_2 =  3.58806 \; .
\end{equation}
\eml

Comparing  with (\ref{EVNUM3})  we see that both codes yield  the
same results within the numerical  accuracy.  This is reassuring,
since, as already stated, the codes were developed  independently
from each other.

\ind
A cubic fit to a total  of 30 data points  produced  with Code~II
yields

\bml
\label{EVNUM5}
\begin{equation}
M_1 = 3.62758 \; , 
\end{equation}
\begin{equation}
M_2 = 3.58260 \; ,
\end{equation}
\begin{equation}
M_3 = 0.06450 \; ,
\end{equation}
\eml

in fair  agreement  with  the variationally  obtained  values  of
Table~\ref{II}.  To check the stability of this fit we can compare with
the extension to fourth order,

\bml
\label{EVNUM6}
\begin{equation}
M_1 = 3.62756 \; , 
\end{equation}
\begin{equation}
M_2 = 3.58314 \;, 
\end{equation}
\begin{equation}
M_3 = 0.06280 \; , 
\end{equation}
\begin{equation}
M_4 = 0.001232 \; .
\end{equation}
\eml

This leads to an estimate  of roughly  $10^{-6}$,  $10^{-4}$  and
$10^{-2}$ for the (absolute) numerical error in the first, second
and third order  coefficient,  respectively.   Furthermore  it is
gratifying  to note that the fourth order coefficient,  $M_4$, is
numerically small.

\ind
As is obvious from the discussion above, the numerical errors for
the 't~Hooft  model are much smaller than those for the Schwinger
model.  The basic reason for this is the Schwinger model anomaly.
Note that in the bound state equation the anomaly factor $\alpha$
multiplies an integral over the wave function.  To evaluate this,
one needs the wave function as a whole. In the 't~Hooft model, on
the other  hand,  this  term  is absent  and the eigenvalues  are
dominated entirely by the endpoint behaviour of the wave function
which  is known exactly.   This makes the numerical  errors  much
smaller (see also Tables~\ref{XIX}, \ref{XX}).

\section{Discussion and Conclusion}
\label{sec:Disc-Concl}

\ind
In  the  preceding  sections  we have  calculated  the  expansion
coefficients $M_i$ in the series

\begin{equation}
M^2 = \alpha + M_1 m + M_2 m^2 + M_3 m^3
\label{MEXP}
\end{equation}

for the `pion' mass squared of both the 't~Hooft ($\alpha=0$) and
the Schwinger  model  ($\alpha=1$).   In order  for the expansion
(\ref{MEXP})  to make sense we have considered  only small masses
$m \ll 1$ {\em i.e.}~$m \ll \mu_0$ in the original units. For the
't~Hooft  model we have $\mu_0^2  = g^2 N_C /2\pi$  with $N_C \to
\infty$, $g^2 N_C$ fixed.  Thus, we are working in the {\em weak}
coupling  phase where the limit $N_C \to \infty$  ($g \to 0$), is
taken  {\em before}  the limit $m \to 0$, or, equivalently,  such
that  one  always   has  $m  \gg  g  \sim  1/\sqrt{N_C}   \to  0$
\cite{Zhi86}  \cite{Zhi96}.   For the Schwinger model, $\mu_0^2 =
e^2 /\pi$, so that small fermion mass corresponds to {\em strong}
coupling.

\ind
We have used analytical, computer algebraic and numerical methods
of  different  accuracy.  Nonetheless,  the  overall  picture  is
intrinsically  consistent.  Our main findings are the following:

(i) The first order coefficient,  $M_1$, in the expansion  of the
`pion' mass squared is independent  of the anomaly $\alpha$, {\em
i.e.}~the  same in both the 't~Hooft  and Schwinger  model.  This
confirms  the results of 't~Hooft \cite{tHo74}  \cite{tHo75}  and
Bergknoff \cite{Ber77}.

(ii) The 't~Hooft-Bergknoff  value, $M_1 = 2\pi /\sqrt{3}$,  is a
`variational invariant': it does not get altered by extending the
space   of  variational   parameters.    This  has  been  checked
analytically and numerically. 

(iii) In the Schwinger  model, the 't~Hooft-Bergknoff  value does
not change upon inclusion  of higher particle sectors.   Only the
second  and  third  order  coefficients,  $M_2$  and  $M_3$,  are
affected.

(iv) Thus, for the Schwinger  model, there remains  a few percent
discrepancy  in the coefficients  $M_1$  and  $M_2$  compared  to
bosonization results.

(v) The variational  calculation  yields the `pion' wave function
to a high accuracy.  The endpoint behaviour is reproduced exactly
(in leading  order  in $m$).   The behaviour  in the intermediate
region,  $0 < x < 1$, gets improved  as can be seen  from  Fig.~3
below.

\ind
In Table \ref{XIX} we summarize  our  optimum  final  results  for the
't~Hooft model ($\alpha$  = 0) and compare with results that have
been obtained previously.

\widetext

\begin{table}
\renewcommand{\arraystretch}{1.5}
\caption{\label{XIX}  The expansion  coefficients  $M_i$  for the
  't~Hooft model ($\alpha$ = 0). The errors of our results obtained
  within 2-particle Tamm-Dancoff (2PTD) approximation are estimated by
  comparing the last two lines in Table~\ref{II} (for the variational
method) and the different polynomial fits (\protect\ref{EVNUM3} -
\protect\ref{EVNUM6}) (for the numerical results).  The numerical inaccuracies
given in the last line were estimated  by us using the polynomial
fit method.}

\vspace{.3cm}

\begin{tabular}[t]{l  l  l  l  } 
& $M_1$ & $M_2$ & $M_3$ \\ \hline
variational 2PTD & $2\pi/\sqrt{3}$ = 3.627599 & 3.581055 $\pm \, 3
\cdot 10^{-6}$ & 0.061793 $\pm \, 3 \cdot 10^{-6}$ \\ 
numerical  2PTD & 3.62758  $\pm \, 2\cdot 10^{-5}$  & 3.5829 $\pm
\, 3 \cdot 10^{-4}$ & 0.064 $\pm \, 1 \cdot 10^{-3}$ \\
't~Hooft \cite{tHo75}  & $ 2\pi/\sqrt{3}$  = 3.627599 & $-$ & $-$
\\
Burkardt  \cite{Bur89a}  & $2\pi/\sqrt{3}$  = 3.627599 & 3.5812 &
$-$ \\ 
Li \cite{Li86} & 3.64 $\pm$ 0.06 & $-$ & $-$ \\
Li et al. \cite{BLW87} & 3.64 $\pm 0.03$ & 3.60 $\pm 0.06$ & 0.04
$\pm 0.04$
\end{tabular} 
\end{table}

\narrowtext

\ind
Upon  inspection  of Table~\ref{XIX}  one  finds  a very  good  overall
consistency  of  the  results.  All  given  values  are  in  good
agreement within error bars.  The scale is set by the variational
results which are the most accurate ones. They are matched by our
numerical results as well as by previous analytical and numerical
calculations.

\ind
As far as the latter are concerned,  a few remarks  are in order.
The  coefficients   $M_i$   displayed   in  the  last   line   of
Table~\ref{XIX} are obtained by performing polynomial fits to the
values calculated  numerically  in \cite{BLW87}  (which basically
agree with those from 't~Hooft's  original calculation  displayed
in Fig.~5 of \cite{tHo74}).   The averaged  result of the fits is
shown including  an estimate  of the errors.   It should  also be
stressed that the data in \cite{BLW87} are mainly obtained for $m
> 0.5$ whereas  the bulk of our values for the $M_i$ are obtained
for $m<0.1$.   Nevertheless,  the agreement  with our results  is
satisfactory within error bars.

\ind
From  Table~\ref{XIX}  it is obvious  that all numerical  results
favour    the   first   order   't~Hooft-Bergknoff    value    of
$2\pi/\sqrt{3}$.   This agreement is particularly  gratifying for
the  data  of  \cite{BLW87}   which  were  not  obtained  via  LC
techniques  but within ordinary  quantization  and even within  a
different  gauge, namely axial gauge.  The agreement is therefore
highly non-trivial \cite{BG78}.

\ind
For  the  Schwinger  model,  ($\alpha$  = 1),  we  summarize  our
findings in Table~\ref{XX}.   For comparison  we have also listed
analytical results obtained via bosonization  techniques and data
which  we  extracted  from  polynomial  fits  to  numerical  LFTD
\cite{MP93},  DLCQ \cite{BEP87}  and lattice results \cite{CH80}.
The  contents  of Table~\ref{XX}  are  graphically  displayed  in
Fig.~2, where we have chosen rescaled units as in \cite{FPV96}.

\widetext

\begin{table}
\renewcommand{\arraystretch}{1.5}
\caption{\label{XX}  The  expansion  coefficients  $M_i$  for the
  Schwinger model ($\alpha$ = 1).  The errors for the variational
  2-particle Tamm-Dancoff (2PTD) approximation are estimated by
  comparing the last two lines in Table~\ref{VI}.  The errors of our
  numerical calculations are obtained from comparing polynomial fits
  of different order to the numerical results of Tables~\ref{XV} and
  \ref{XVI}.  Errors of other results are given where they could be
  estimated analogously.}

\vspace{.3cm} \begin{tabular}[t]{l l l l } 
& $M_1$ & $M_2$ & $M_3$ \\ \hline 
variational 2PTD & $2\pi/\sqrt{3}$  = 3.627599 & 3.308608 $\pm \,
4 \cdot 10^{-6}$ & 0.348204 $\pm 1 \, \cdot 10^{-6}$ \\
numerical 2PTD & 3.6268 $\pm 8 \cdot 10^{-4}$ & 3.32 $\pm 0.02 $ &
0.28 $\pm 0.06 $ \\
numerical 6PTD & 3.6267 $\pm 4 \cdot 10^{-4}$ & 3.22 $\pm 0.02 $ &
0.5 $\pm 0.1 $ \\
numerical  4PTD  \cite{MP93}  & 3.62 $\pm 0.07$  & $3.2 \pm 0.3$ &
$0.3 \pm 0.2 $ \\
DLCQ \cite{BEP87}  & 3.7 $\pm 0.2 $ & 3.5 $\pm 0.3 $ & $-$ \\
lattice \cite{CH80} & 3.5 $\pm 0.2 $ & 3.7  & 0.02 \\
Adam \cite{Ada96} & $2 e^\gamma$ = 3.562146 & 3.3874 & $-$ \\
Fields et al. \cite{FPV96} & $2 e^\gamma$ = 3.562146 & 3.387399 &
$-$ \\
\end{tabular} 
\end{table}

\narrowtext

\begin{figure}
\begin{center}
\caption{The  rescaled `pion' mass $\bar M = M/\protect\sqrt{1  +
    m^2}$ as a function of the fermion mass $m$.  The short-dashed
  curve represents a `phenomenological' parametrization for $\bar
  M$ with $M^2 = 1 + M_1 m + 4 m^2$, which becomes exact for very
  small and very large $m$ and thus smoothly interpolates between the
  strong and weak coupling regions.  The long-dashed curve is our
  second-order and the solid curve our third-order result.  As
  expected, mass perturbation theory breaks down as $m$ becomes of
  order 1.  The crosses are the lattice results \protect\cite{CH80},
  the diamonds the lattice results \protect\cite{CK86}, which go down
  to comparatively small masses, however with large errors.  The
  circles are LFTD results \protect\cite{MP93}, whereas the squares
  \protect\cite{BEP87} and triangles \protect\cite{Els94}, as quoted
  in \protect\cite{FPV96}, are DLCQ results.  The values corresponding
  to the triangles are {\em not} included in Table~\ref{XX}.  The 2\%
  discrepancy in $M_1$ is invisible within the resolution of the
  figure.}  \vspace{0.5cm}
\includegraphics[width=\graphicwidth]{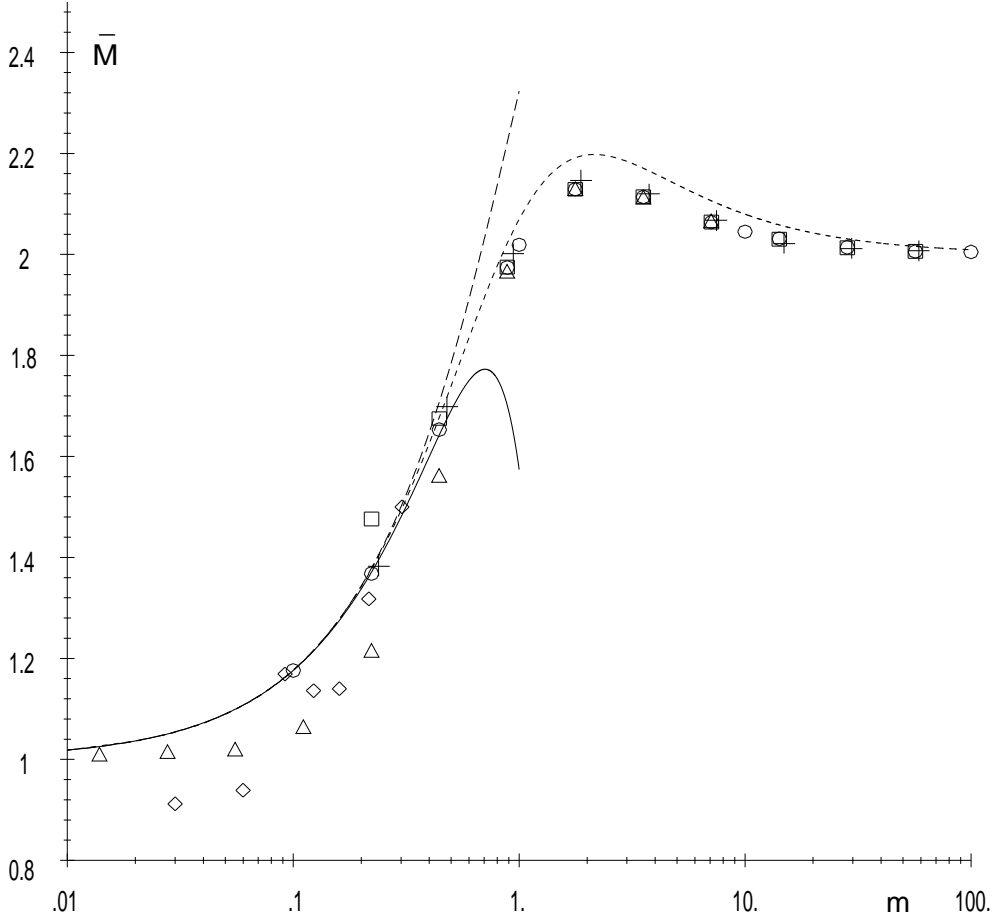}
\end{center}
\end{figure}

\ind
As already  stated, due to the anomaly  the numerical  errors for
the Schwinger  model  are at least an order  of magnitude  larger
than for the 't~Hooft  model.   In addition,  the 2\% discrepancy
between our LC results and the analytic bosonization results does
not get resolved.  However, we have shown that the discrepancy is
not due to (i) inaccuracies  in the wave function or (ii) neglect
of contributions  from the next higher Fock sectors (four and six
particles).   It should be mentioned that other numerical methods
(DLCQ, lattice) are by far too inaccurate to distinguish  between
the 't~Hooft-Bergknoff  and the bosonization value.  We will come
back to these issues in a moment.

\ind
As is well known, \cite{FPV96} \cite{MP93} \cite{vdS96}, the DLCQ
data  are comparatively  inaccurate  for small  fermion  mass $m$
({\em  i.e.}~large  coupling).   This is due to the fact that the
dominating   feature  of  the  LC  wave  function,  its  endpoint
behaviour,   is  not  very   accurately   reproduced   using   an
equally-spaced  momentum grid.  The poor convergence  of DLCQ for
small $m$ has recently been overcome  by incorporating  the exact
endpoint behaviour \cite{vdS96}.

\ind
The lattice data generally suffer form the same disease of having
rather large error bars for small $m$. The chiral limit, $m = 0$,
can only be reached via extrapolation in a very inaccurate manner
unless  one uses the Schwinger  result  $M^2 (m=0) = 1$ as a bias
\cite{CH80}.  A representative  collection of lattice results for
$M_1$ is given in Table~\ref{XXI}.  For the convenience of the reader
we also list the condensate $\cond = - M_1 /4\pi$.

\begin{table}
\renewcommand{\arraystretch}{1.5}
\caption{\label{XXI} Lattice   results   for  the  first   order
coefficient $M_1$ and the (negative of the) condensate, $-\cond
= M_1 / 4\pi$.  The results are quoted in chronological  order.
They should be compared with the bosonization  results,  $M_1 =
3.562$,  $-\cond  = 0.283$  and  the  't~Hooft-Bergknoff  values,
$M_1 = 3.628$, $-\cond = 0.289$.}

\vspace{.3cm} 

\begin{tabular}[t]{c c c c c c c} 
&  BKS76\tablenote[1]{Banks,   Kogut,  Susskind  \cite{BKS76}}  &
CKS76\tablenote[2]{Carroll,     Kogut,     Sinclair,     Susskind
\cite{CKS76}} & CH80\tablenote[3]{Crewther,  Hamer \cite{CH80}} &
MPR81\tablenote[4]{Marinari,   Parisi,   Rebbi  \cite{MPR81}}   &
HKC82\tablenote[5]{Hamer,      Kogut,     Crewther,     Mazzolini
\cite{CHK82}} & CK86\tablenote[6]{Carson,  Kenway \cite{CK86}} \\
\hline
$M_1$ & 3.42 & 3.644 & 3.48 & 2.97 & 3.33 & 3.77 \\
$-\cond$  &  0.27  &  0.290  &  0.28  & 0.24  & 0.26 & 0.30
\end{tabular} 
\end{table}

\ind
The values for the Schwinger model condensate in Table~\ref{XXI} should
be  compared  with  our  results.  If  we  assume  that  formulae
(\ref{COND})  and (\ref{ADAM1})  are directly  applicable  to our
calculation we get

\begin{eqnarray}
\cond  &=&  -  \frac{M_1}{4\pi}  =  -  \frac{1}{2\sqrt{3}}  = -
0.28868 \; , \label{COND1} \\
\cond &=& - \frac{1}{2\pi}  \sqrt{\frac{M_2}{A+B}}  = - 0.28015
\; , \label{COND2}
\end{eqnarray}

while the standard result is

\begin{equation}
\cond = - \frac{e^\gamma}{2\pi} = - 0.28347 \; .
\label{COND_BOS}
\end{equation}

As  already  mentioned  in  the  introduction,  the  few  percent
discrepancies   in  $M_1$  and  $M_2$  immediately   affect   the
condensate.   Note that the condensate  value  obtained  from the
second  order coefficient  is closer to the bosonization  result,
the relative  error being 1.2 \% compared  to the 1.8 \% at first
order.   Furthermore,  the  exact  result  (\ref{COND_BOS})  lies
{\em between}  our first and second  order values,  (\ref{COND1})
and (\ref{COND2}),  so that their mean value, $\cond = -0.28442$,
is much closer to (\ref{COND_BOS}), the error being only 0.3\%. A
similar reduction  of errors is actually at work if one considers
the `pion' mass-squared  of the Schwinger  model as a function of
$m$.  The first order overshoots the bosonization value while the
second  order contribution  is too small.   Adding  both one gets
closer  to the exact result,  at least if the mass $m$ is not too
tiny.   For the value  of Fig.~1,  $m$ = 0.1, one finds to second
order $M^2 (0.1) = 1.39585$, while $M^2_{\mbox{\scriptsize{Bos}}}
(0.1) = 1.39009$. The relative difference is only 0.4\%. Possible
resolutions   of   the   discrepancies   between   (\ref{COND1}),
(\ref{COND2})  and (\ref{COND_BOS})  will be discussed at the end
of this section.

\ind Apart from the `pion' mass squared, the eigenvalue in the
't~Hooft-Bergknoff equation, we have also calculated the associated
eigenfunction, the light-cone wave function, with high accuracy.  The
latter has been obtained with high accuracy as can be seen from the
good convergence of the alternatively defined mass squared
coefficients, $\tilde M_i$ towards the variational estimates, $M_i$
(see Section~\ref{sec:High-Order-Extension}).  If we denote $\phi_0
(x) = x(1-x)$, our most accurate variational ansatz for the wave
function can be written as

\begin{equation}
\phi  \Big[\beta,  a,  b,  c,  d  \Big]  =  \phi_0^\beta  +  a \,
\phi_0^{\beta+1}  + b \, \phi_0^{\beta+2} + c \, \phi_0^{\beta+3}
+ d \,  \phi_0^{\beta+4} \; . 
\end{equation}

According   to  our  mass  perturbation   theory,   each  of  the
variational  parameters is expanded in powers of the fermion mass
$m$,  the  coefficients   being  denoted  $\beta_1$,   $\beta_2$,
$\beta_3$, $a_1$, $a_2$, etc.  From 't~Hooft's endpoint analysis,
$\beta$ is known exactly.  It turns out to be independent  of the
anomaly  ($\alpha$)  and thus  is the same  for the 't~Hooft  and
Schwinger  model.  In Table~\ref{XXII}  we compare  the exact expansion
coefficients  of $\beta$ with their variational  estimates.   The
best estimates for the other variational parameters are listed in
Table~\ref{XXIII} for both the 't~Hooft and Schwinger model.

\widetext

\begin{table}
\renewcommand{\arraystretch}{1.5}
\caption{\label{XXII} Comparison  of the exact expansion  coefficients  of the
  endpoint exponent $\beta$ with their best variational estimates for
  both the 't~Hooft ($\alpha$ = 0) and the Schwinger model ($\alpha$ =
  1).}

\vspace{.3cm} 

\begin{tabular}[t]{l d d d }
& $\beta_1$ & $\beta_2$ & $\beta_3$ \\ \hline  
exact  &  $\sqrt{3}/\pi$  =  0.55132890  &  0  &  $-\beta_1/10  =
-$0.05513289 \\
variational ($\alpha$ = 0) & $\sqrt{3}/\pi$  = 0.55132890 & 0.00051
& $-$0.053 \\
variational ($\alpha$ = 1) & $\sqrt{3}/\pi$  = 0.55132890 & 0.00063
& $-$0.053 
\end{tabular} 
\end{table}

\narrowtext

\widetext

\begin{table}
\renewcommand{\arraystretch}{1.5}
\caption{\label{XXIII}   Best   estimates   for   the  variational
parameters  in  the  LC  wave  function  for  both  the  't~Hooft
($\alpha$ = 0) and Schwinger model ($\alpha$ = 1).}

\vspace{.3cm} 

\begin{tabular}[t]{l c c c c c c d c }
& $a_1$  & $a_2$ & $b_1$ & $b_2$  & $c_1$ & $c_2$ & $d_1$ & $d_2$
\\ \hline
$\alpha = 0$ & 0.92 & 1.19 & $-2.2$ & $-3.0$ & 7.0 & 9.2 & $-$9.9
& $-11.8$ \\
$\alpha  = 1$ & 1.42  & 1.03  & $-2.7$  & $-3.1$  & 8.8  & 9.6  &
$-$12.5 & $-12.1$
\end{tabular} 
\end{table}

\narrowtext

\ind
The  errors  in the  coefficients  are  comparatively  large  and
growing from the left to the right in Table~\ref{XXIII}. Good numerical
convergence is evident for the coefficients of $a$. For the other
coefficients  our method does not provide enough iteration  steps
to make convergence  explicit.   However, the sensitivity  of the
mass expansion coefficients  $M_i$ and the wave functions  on the
parameters $b$, $c$, $d$ is very weak (see Fig.~3).

\begin{figure}[tb]
\begin{center}
\caption{The light-cone wave function of the 't~Hooft model `pion'
for  $m$  = 0.1.   The solid  curve  represents  the result  from
't~Hooft's  original  ansatz,  the dashed  curve  our best result
(with maximum  number of variational  parameters).   At the given
resolution,  however,  the  curves  of {\em  all}  extensions  of
't~Hooft's  ansatz  (a,  b,  c,  d) lie  on top  of each  other.}
\vspace{0.5cm}
\includegraphics[width=\graphicwidth]{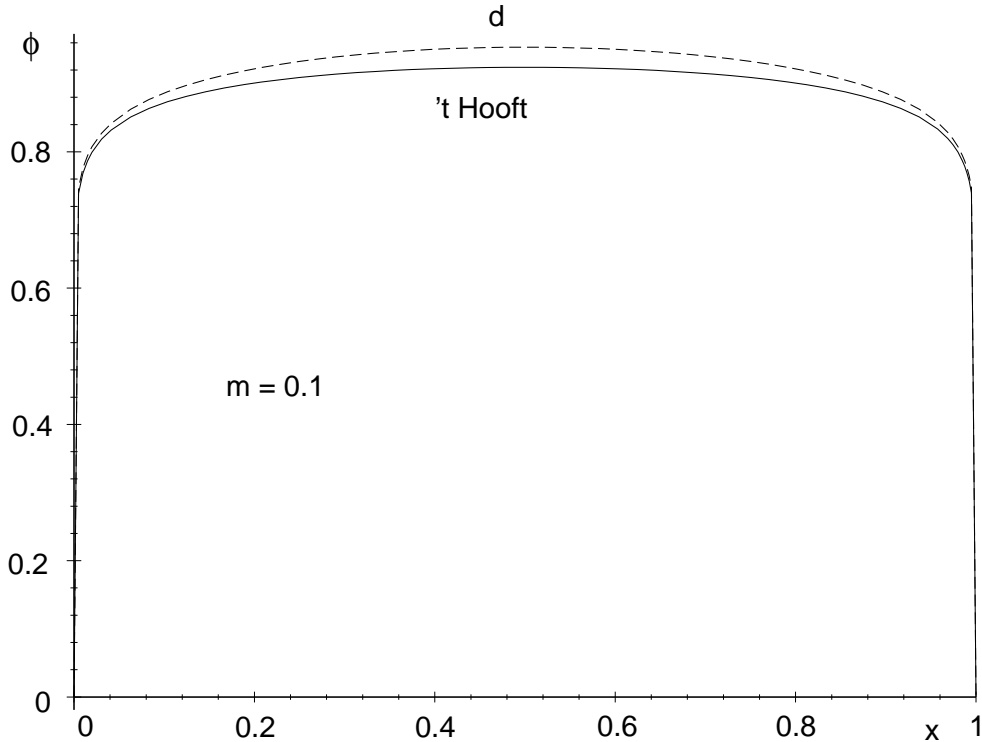}
\end{center}
\end{figure}

\ind
Let us finally investigate  possible  sources for the few-percent
discrepancies  in  the  Schwinger  model  results  like  for  the
condensate   values,   (\ref{COND1})   and   (\ref{COND2})   {\em
vs.}~(\ref{COND_BOS}).   First  of all, it has to be reemphasized
that  the  light-cone   calculations   are  conceptually   rather
different  from the usual treatment based on bosonization  within
standard  equal-time  quantization.   We are studying  the lowest
lying meson as a relativistic  fermion-anti-fermion  bound state.
The fermionic degrees of freedom are {\it explicitly}  taken into
account as they define our Fock basis.  In the bosonized  theory,
which  is a sine-Gordon  model  with the Lagrangian  \cite{CJS75}
\cite{Col76}

\begin{equation}
{\cal  L} = \frac{1}{2}  \partial_\mu  \phi  \partial^\mu  \phi -
\frac{1}{2} \mu_0^2 \phi^2 - c m \mu_0 \cos \sqrt{4\pi} \phi \; ,
\label{SINE_GORDON}
\end{equation}

the `pion'  is the elementary  particle  described  by the scalar
field $\phi$.  One calculates its mass by perturbation  theory in
the  `coupling'  $c  m \mu_0$,  $c = e^\gamma  / 2\pi$.   In  the
bosonized  theory, this `coupling'  is actually dependent  on the
normal-ordering  scale  used  to  renormalize  the  theory.   The
standard  choice  is the most natural  one, namely  the Schwinger
boson  mass $\mu_0$, which explicitly  appears  in the Lagrangian
(\ref{SINE_GORDON}).  It is, however,  not very hard to determine
the   general   scale   dependence   of   quantities   calculated
within   perturbation   theory.    All  one  needs  is  Coleman's
re-normal-ordering  prescription  \cite{Col75} which for the case
at hand is

\begin{equation}
N_{\mu_0}  \cos \sqrt{4\pi}  \phi = \frac{\mu}{\mu_0}  N_\mu \cos
\sqrt{4\pi} \phi \; .
\label{RNO}
\end{equation}

Here, $N_{\mu_0}$ and $N_\mu$ denote normal-ordering with respect
to $\mu_0$ and $\mu$, respectively.  From (\ref{RNO}) one derives
the following identity for the boson mass squared (setting $\mu_0
= 1$),

\begin{eqnarray}
M^2  &=&  1 + M_1 \, m + M_2 \, m^2 + M_3 \, m^3 + \ldots  \nn \\
&=& 1 + M_1^\prime  \, \mu m + M_2^\prime  \, \mu^2 m^2 +
M_3^\prime \, \mu^3 m^3 + \ldots \; . \label{SCALE} 
\end{eqnarray}

In this expression, the $M_i$ denote the coefficients  calculated
with  the  standard  normal-ordering  scale  $\mu_0$,  where  all
tadpoles  can  be  set  to  zero,  while  the  $M^\prime_i$   are
calculated  with  normal-ordering  scale  $\mu$,  where  one  has
non-vanishing  tadpole  contributions.   In order that the `pion'
mass   be  scale   independent   one  has  to  have   that  (upon
reintroducing $\mu_0$)

\begin{equation}
M_n^\prime = \left( \frac{\mu_0}{\mu} \right)^n M_n \; .
\label{RESCALE}
\end{equation}

We  have  checked  this  identity  explicitly  in the  first  two
non-trivial   orders   of  mass  perturbation   theory  with  the
Lagrangian (\ref{SINE_GORDON}).   It is also interesting  to note
that the scale dependence of the coefficients  might equivalently
be associated with a multiplicative rescaling of the fermion mass
$m$ by defining  $m^\prime  = \mu m$ and using $m^\prime$  as the
expansion parameter in (\ref{SCALE}).

\ind
Now, the upshot of all this is the following.   As we are working
entirely  in the  fermionic  representation,  we do not  a priori
know,  to which  normal-ordering  scale  of the  {\em  bosonized}
theory our results  correspond.   If we assume that it is not the
Schwinger  mass $\mu_0$ but a different scale $\mu$, we can match
the first order coefficients by choosing

\begin{equation}
\mu      =      \frac{M_1}{M_1^\prime}       \mu_0      =
\frac{e^\gamma}{\pi/\sqrt{3}} \mu_0 \; ,
\label{NEWSCALE}
\end{equation}

where $M_1^\prime$  now denotes our light-front  result.  Quite a
similar point of view has been taken by Burkardt  in his paper on
light-cone  quantization  of the sine-Gordon  model \cite{Bur93}.
However, as is obvious from (\ref{RESCALE}),  the change of scale
(\ref{NEWSCALE})    propagates    systematically    through   all
coefficients.   In particular, if $M_1^\prime$ is larger than the
corresponding  bosonization  result,  the same has to be true for
all $M_i^\prime$.  Our coefficient $M^\prime_2$, however, already
fails this requirement.   We thus conclude  that a possible scale
dependence alone cannot explain the discrepancy.

\ind
An additional  source of error could still be contributions  from
higher Fock sectors. We have shown numerically that the first few
low orders have very little impact on the expansion  coefficients
$M_i$.   The contribution  from each Fock sector  might  thus  be
exponentially  small.  The whole series, however,  if it could be
summed  up, might  add something  seizable.   Let us assume,  for
example,  that the contribution  of higher Fock sectors  to $M_2$
has  the  following   form\footnote{We   thank  M.~Burkardt   for
suggesting the following example.}

\begin{equation}
\delta M_2 = \sum_{n=2}^\infty c_{2n}(m) \; ,
\label{SERIES}
\end{equation}

where $2n$ is the label of the $2n$-particle  sector.   If we had
$c_{2n}  (m) = e^{-2nm}$,  summation  of the series  would  yield
$\delta M_2 = 1/2m + O(1)$ and thus effectively a contribution to
the first  order  in $m$.   As we are not able  to calculate  the
real high order  behaviour  of the series  in (\ref{SERIES}),  we
cannot,  at the moment,  make  any definite  conclusion  about  a
possible relevance of such summation effects.  All we can say is,
that if such effects are present, they must be numerically small,
{\em i.e.}~at the few percent level.

\ind
As a first step towards  getting  some better  control  of higher
particle sectors one would like to develop some tools to estimate
their contributions analytically.   This amounts to determine the
coefficients $c_{2n}(m)$, at least for small $n$.  So far, higher
Fock states  have only been included  numerically.   

\ind
It might also be interesting to calculate excited states. For the
Schwinger model, this again requires the inclusion of higher Fock
components  as pointed out by Mo and Perry \cite{MP93}.   For the
't~Hooft   model,  on  the  other  hand,  one  can  stay  in  the
two-particle sector. One might then try to modify the variational
procedure, basically by taking into account the growing number of
nodes,  with the aim to match  the low energy  spectrum  with the
high  energy  behaviour  derived  by  't~Hooft.   Work  in  these
directions is underway.

\section*{Acknowledgements}
\label{sec:Acknowledgements}

It is a pleasure to thank T.~Sugihara for providing the numerical
data  yielding   the  fits  (\ref{EVNUM5})   and  (\ref{EVNUM6}).
K.H.~and  T.H.~would  like  to  express  their  gratitude  to the
organizers  of  the  1997  Les  Houches  workshop  on `Light-Cone
Quantization  and Non-Perturbative  Dynamics',  in particular  to
P.~Grang\'e, for providing such a stimulating atmosphere in which
this work was finished.  They thank the participants  M.~Burkardt
and S.~Pinsky for valuable discussions.  T.H.~is also indebted to
C.~Adam,  F.~Lenz,  L.~Lipatov,  A.~Neveu and B.~van de Sande for
useful hints and suggestions.  T.H.~and C.S.~thank S.~Simb\"urger
for  his  assistance  with  some  mathematical   intricacies  and
E.~Werner    for    continuous    support    and   encouragement.
K.H.~acknowledges  financial  support  provided  by the  Sumitomo
foundation (No.~960517).

\end{document}